\documentclass[twocolumn]{aa}
\usepackage{graphicx}
\usepackage{txfonts}
\begin{document}
%
   \title{CLOUDS search for variability in brown dwarf atmospheres.}
   \titlerunning{Infrared spectroscopic time series of L/T transition brown dwarfs}

   \subtitle{Infrared spectroscopic time series of L/T transition brown dwarfs\thanks{Based on observations obtained at the European Observatory, Paranal, Chile,
             under programme 71.C-0559.}}

   \author{
       B. Goldman\inst{1,2}
          \and
	  M.~C.~Cushing\inst{3}\thanks{Visiting Astronomer at the Infrared Telescope Facility, which is operated by the University of Hawaii under Cooperative Agreement no. NCC 5-538 with NASA, Office of Space Science, Planetary Astronomy Program.} 
          \and
          M.~S.~Marley\inst{4}
          \and
          \'E.~Artigau\inst{5}
          \and
          K.~S.~Baliyan\inst{6}
          \and
\\        V.~J.~S.~B\'ejar\inst{7}
          \and
        J.~A.~Caballero\inst{2,8} 
          \and
          N.~Chanover\inst{1}
          \and
          M.~Connelley\inst{9}
          \and
          R.~Doyon\inst{10}
          \and
          T.~Forveille\inst{11,12}
          \and
\\          S.~Ganesh\inst{6}
          \and
        C.~R.~Gelino\inst{1,13}
          \and
          H.~B.~Hammel\inst{14}
          \and
          J.~Holtzman\inst{1}
          \and
          S.~Joshi\inst{15}
          \and
          U.~C.~Joshi\inst{6}
          \and
\\         S.~K.~Leggett\inst{16}
          \and
        M.~C.~Liu\inst{9}
          \and
          E.~L.~Mart\'\i n\inst{8}
          \and
          V.~Mohan\inst{17}
          \and
          D.~Nadeau\inst{10}
          \and
          R.~Sagar\inst{15}
          \and
          D.~Stephens\inst{18}
          }

   \offprints{B.~Goldman, {\tt go{\,\hspace{-1pt}}ld\hspace{-1pt}\,man{\,}@mp{}ia.de}}

   \institute{
              Department of Astronomy, New Mexico State University, 
              Las Cruces, NM 88003,
              U.S.A.
         \and 
              M.P.I.A., K\"onigstuhl~17, 69117 Heidelberg, Germany
         \and 
                  Steward Observatory, University of Arizona, Tucson, AZ 85721, 
              U.S.A.
          \and 
              NASA Ames Research Center, Moffett Field, CA 94035,
              U.S.A.
         \and 
              Gemini Observatory, Southern Operations Center, A.U.R.A., Inc., Casilla 603, La Serena, Chile
         \and 
              Astronomy \& Astrophysics Division, Physical Research Laboratory, Navarangpura, Ahmedabad 380 009, India
          \and 
             GTC project. Instituto de Astrof{\'\i}sica de Canarias, E-38205 La~Laguna, Tenerife, Spain
         \and 
              Instituto de Astrof{\'\i}sica de Canarias, E-38205 La~Laguna, Tenerife, Spain
         \and 
              Institute for Astronomy, University of Hawaii, 
              2680 Woodlawn Drive, Honolulu, HI 96822,
              U.S.A.
          \and 
              Observatoire du Mont M\'egantic et D\'epartement de Physique, Universit\'e de Montr\'eal, H3C 3J7, C.P. 6128, Montr\'eal, 
              Canada
         \and 
              Canada-France-Hawaii Telescope Corporation,
              65-1238 Mamalahoa Highway,
              Kamuela, HI 96743, Hawaii, 
              U.S.A.
         \and 
              Observatoire de Grenoble,
              414 rue de la Piscine,
              Domaine Universitaire de S$^{\mathrm t}$ Martin d'H\`eres,
              F-38041 Grenoble,
              France
         \and 
              Spitzer Science Center, MC 220-6, California Institute of Technology, Pasadena, CA 91125,
              U.S.A.
         \and 
              Space Science Institute, 4750 Walnut Street, Suite 205, Boulder, CO, 80301,
              U.S.A.
         \and 
                 Aryabhatta Research Institute of Observational Sciences (ARIES), 
                 Manora Peak, Nainital--263129, Uttaranchal,
                 India
         \and 
                Joint Astronomy Centre, 660 North A'ohoku Place, Hilo, HI 96720,
                U.S.A.
         \and 
                 Inter-University Centre for Astronomy and Astrophysics (IUCAA), 
                 Post Bag 4, Ganeshkhind, Pune 411007, India
         \and 
                Department of Physics and Astronomy, Brigham Young University, Provo, UT 84602 
                U.S.A.
       }

   \date{Received; accepted}

   \abstract
   {L-type ultra-cool dwarfs and brown dwarfs have cloudy atmospheres that could host weather-like phenomena.  The detection of photometric or spectral variability would provide insight into unresolved atmospheric {heterogeneities}, such as holes in a global cloud deck. Indeed, a number of ultra-cool dwarfs have indeed been reported to vary.
Additional time-resolved spectral observations of brown dwarfs offer the opportunity to further constrain and characterize atmospheric variability.
   }
   {It has been proposed that growth of {heterogeneities} in the global cloud deck may account for the L- to T-type transition as brown dwarf photospheres evolve from cloudy to clear conditions.  Such a mechanism is compatible with variability.
   We searched for variability in the spectra of five L6 to T6 brown dwarfs in order to test this hypothesis. } 
   {We obtained spectroscopic time series
   using the near-infrared spectrographs {ISAAC} on VLT--ANTU, over 0.99--1.13\,$\mu$m, 
   and {SpeX} on the Infrared Telescope Facility for two of our targets, in $J, H$ and $K$ bands. 
   {We search for statistically variable lines and correlation between those.}\,
   }
   {High spectral-frequency variations are seen in some objects, but these detections are marginal and
   need to be confirmed.
   We find no evidence for large amplitude variations in spectral morphology and we place firm upper limits
   of {2 to 3}\,\% on broad-band variability, depending on the targets and wavelengths, on the time scale of a few hours.
   In contrast to the rest of the sample, the T2 transition brown dwarf {SDSS\,J1254$-$0122} shows numerous variable features, but a secure variability diagnosis would require further observations.}
  {Assuming that any variability arises from the rotation of patterns of large-scale clear and cloudy regions across the surface,
   we find that the typical physical scale of cloud cover disruption should be smaller than 5--8\,\% of the disk {area} for four of our targets, using simplistic heterogeneous atmospheric models.
   {The possible variations seen in  {SDSS\,J1254$-$0122} are not strong enough to allow us to confirm the cloud breaking hypothesis.}\,
   }
  
  
   \maketitle
%

\section{Introduction}

  L/T transition brown dwarfs are an informal class of sub-stellar objects comprising the latest L-type
  and the earliest T-type dwarfs (roughly {L8 to T4}). 
  Their near-infrared spectra may exhibit both CO and CH$_4$ absorption and 
  they have $J-K$ colours intermediate between those of the red late-type L dwarfs and the blue mid-T dwarfs (e.g. $0<J-K<1.2$).  
The spectra of L~dwarfs are characterized
by absorption due to metal hydrides (FeH, CrH) and {strong} alkali (K, Na, Rb, Cs) lines,
absorption bands of water and CO, and
the grey opacity of a cloud deck
(Kirkpatrick et~al.~\cite{Kir99}, Mart\'\i n et~al.~\cite{Mar99}).
These clouds are made of condensates of refractory elements such as
Al, Ca, Fe, Mg, Si,
as well as {Ti and V} whose oxides dominate the optical spectral morphology of M dwarfs,
 but which are depleted from cooler atmospheres as they {condense} into dust grains.
The vertical distribution of the clouds is a complex issue,
but  a balance between sedimentation (the equivalent of rain in the case
of water) and upward mixing likely controls their vertical profile (e.g., Ackerman \& Marley~\cite{Ack01}). 
As the atmosphere becomes cooler in late-type L~dwarfs, the grey clouds gradually settle
below the photosphere and 
the emergent spectrum 
becomes more strongly affected by molecular band opacities.
In the next cooler spectral class, T, CO {completes its reduction} to CH$_4$ (Burgasser et~al.~\cite{Bur02a}; Geballe et~al.~\cite{Geb02}).
The latter produces strong IR absorption, along with water and molecular hydrogen (collision-induced absorption).
Synthetic spectra predicted by 
atmosphere models 
with a discrete cloud deck reproduce well the colours (Ackerman \& Marley \cite{Ack01}, Burgasser et~al.~\cite{Bur02b}),
the 
near-infrared spectra (Cushing et~al.~\cite{Cus05}), and the 
mid-infrared spectra (Roellig et~al.~\cite{Roe04}) of the L dwarfs.  
Likewise models with no grain opacity {fit well} the spectra of mid- to late-type T~dwarfs (Cushing et~al.~\cite{Cus08}).

However the rapidity of the transition between the cloudy L~dwarf atmospheres
and the clear T~dwarf atmospheres is not well understood.  Proposed mechanisms have included
horizontal fragmentation of the global cloud deck (Burgasser et~al.~\cite{Bur02b}), a rapid increase
in the cloud particle sedimentation efficiency (Knapp et~al.~\cite{Kna04}), or the sinking of a very thin cloud
layer (Tsuji et~al.~\cite{Tsu04}). The case for a dynamic mechanism, as opposed to 
a 
gradual sinking of a uniform cloud deck, was strengthened by 
recent measurements of geometric parallaxes for a significant number of brown dwarfs
(Vrba et~al.~\cite{Vrb04}), 
and the observations of brown dwarfs over a larger wavelength range of their spectra,
particularly in the mid-infrared (Golimowski et~al.~\cite{Gol04}).  
Effective temperatures derived 
independently of atmospheric modeling assumptions 
reveal that the L/T transition takes place over a small effective temperature range (about 200\,K), 
{as was qualitatively first noticed by Kirkpatrick et~al.~(\cite{Kir99}).
This} is difficult to reconcile with a gradual settling of the clouds.
It is also marked by a brightening in the $J$~band with increasing spectral type
({Dahn et~al.~\cite{Dah02};} Knapp et~al.~\cite{Kna04}, their Fig.~8 and 9).
 Burgasser et~al.~(\cite{Bur02b}) interpreted this rapid brightening, and the corresponding $J-K$ colour change, 
as an indication of the fragmentation of the global cloud cover 
{in the L/T transition brown dwarfs' atmospheres.}

One possible way to test the horizontal fragmentation hypothesis would be to search
for variability over a wide range of spectral types.
Large-scale, patchy clouds, like those on Jupiter, would be expected to create photometric variations (Gelino \& Marley~\cite{Gel00}) as the structures rotate across the disk.
If the signature of such
phenomena is preferentially observed in the L/T transition objects this would strengthen the case for cloud fragmentation.
If the typical length scale of horizontal cloud patches are small, however, little to no photometric or spectral variation would be expected. 
{This would also be the case for stable stripes along the brow dwarf parallels.}
Thus detection of variability supports the cloud fragmentation hypothesis, but non-detection cannot rule it out.
{Here we do not put forward any hypotheses on how such heterogeneities could arise.  We aim simply to  test if the surface is heterogeneous on large spatial scales. }

{Other explanations for the L/T transition enigma have been suggested.}
Since we started this project, a number of L/T brown dwarfs have been resolved into binaries composed by a late-L {or early-T} primary, and a fainter T companion (e.g. {Reid et~al.~\cite{Rei01}}, Burgasser et~al.~\cite{Bur05}, Liu et~al.~\cite{Liu06}).
{Although the statistics are still not conclusive}, the seemingly higher rate of binaries among L/T~transition brown dwarfs compared to other brown dwarfs has prompted {Burgasser et~al.~(\cite{Bur05,Bur06b}) and Liu et~al.~\cite{Liu06})} to speculate that binarity could explain the peculiar relation {between effective temperature and spectral energy distribution} observed over the transition.
{In the case where most L/T transition objects would be composed of earlier- and later-type components, we would not expect to find more variability among transition objects than among the populations from which the components are drawn, taking into account the binary effects.
Similarly, our data do not allow to search confirmation of any mechanism that would not produce surface heterogeneities, including uniform and gradual sinking, or thinning (e.g., Knapp et~al.~\cite{Kna04}, Tsuji et~al.~\cite{Tsu04}) of the clouds.}

There have been a number of reports of brown dwarfs exhibiting variability.   
Tinney \& Tolley~(\cite{Tin99}), 
Bailer-Jones \& Mundt~(\cite{CBJ99,CBJ01}), 
Bailer-Jones \& Lamm~(\cite{CBJ03}), 
Clarke et~al.~(\cite{Cla02a})
and
Koen~(\cite{Koe03}) 
studied samples of early to mid L~dwarfs in the $I$~band and reported variability detections  for a fraction of their targets.
Gelino et~al.~(\cite{Gel02}) conducted optical and near-infrared photometry follow-up of a dozen of brown dwarfs{, and found several variations in their targets' light curves}.
{Enoch et~al.~(\cite{Eno03}) conducted a $K_s$-only monitoring program of nine L and T~dwarfs, including three L/T transition dwarfs.
At the 2-$\sigma$ level, they found three variable dwarfs, including one T1 dwarf with a possible 2.96-h period, at the 89\,\% confidence level (C.L.), and did not  find a correlation between variability and spectral type.
Their average detection limit of sinusoidal variations is 12\,\% (99\,\% C.L.).
}
{Clarke et~al.~(\cite{Cla02b}) reported {that Kelu-1 (L3, {Geballe et~al.~\cite{Geb02}}) varies} photometrically in their 2000 optical observations, with a 1.8\,h period,
but this was not observed again in 2002 (Clarke et~al.~\cite{Cla03}).}
{Bailer-Jones \& Lamm~(\cite{CBJ03}) studied the effect of second-order extinction residuals in photometric variability observations.
These residuals are
due to the variable, wavelength-dependent {telluric} absorption of H$_2$O, CO$_2$ and O$_3$
and to the very different colours of the brown dwarfs and the comparison stars.
They estimate these residuals to be of the percent level, and cannot generally be removed.
Our spectroscopic observations are mostly not affected by this effect.
Only deep, non-resolved lines in both the target spectrum {and the telluric absorption spectrum} could be affected.}

H$\alpha$ has been monitored in two L3 dwarfs (Hall~\cite{Hal02}, Clarke et~al.~\cite{Cla03}) and an L5 dwarf (Liebert et~al.~\cite{Lie03}), 
and the emission was found variable in all three objects.
{Burgasser et~al.~(\cite{Bur02c}) studied the peculiar, H$\alpha$ emitting T6.5 brown dwarf 2MASS~J12373919+6526148 using optical spectroscopy and $J$-band photometry. 
A {tentative} shift in the H$\alpha$ line was detected but no photometric variability, at the $\approx 2.5$\,\% level.
A series of eleven 20-min spectra of the L8 dwarf Gl~584C revealed ``subtle'' qualitative variations in the 9250--9600\,\AA\  water absorption band, although a telluric origin cannot be excluded, and redward of the CrH 8600\,\AA\  bandheaad (Kirkpatrick et~al.~\cite{Kir01}). } 
Near-infrared spectroscopic data are more scarce.
Nakajima et~al.~(\cite{Nak00}) cautiously reported a variation in the water band at 1.53--1.58\,$\mu$m 
in the spectra of the T6 {SDSS\,J1624+0029}, 
using the Subaru telescope, although over only 80\,min of time.
{The work most similar to this has recently been published by Bailer-Jones\,(\cite{CBJ08}). 
Two out of four ultracool dwarfs have been found to show correlated spectral numerous variations with 13-nm resolution, often (but not always) associated with one or more known features, particularly water and methane. }

Finally, Morales-Calder{\'o}n et~al.~(\cite{Mor06})  searched for photometric variability in three late L dwarfs  (DENIS-P~J0255$-$4700, 2MASS~J0908+5032, and 2MASS~J2244+2043) with the IRAC photometer on {\em Spitzer Space Telescope}. Two out of the three objects studied exhibited some slight modulation in their light curves at 4.5, but not $8.8\,\rm \mu m$. 

In short, despite numerous variability searches with a great variety of instruments and technique, there has yet been no definitive detection of luminosity variability in a {field} brown dwarf.
Here we report on a search for spectroscopic variability in a selection of L/T transition objects 
by the {\sc Clouds} collaboration (Continuous Longitude Observations of Ultra-cool Dwarfs, Goldman et~al.~\cite{Gol03})\footnote{\tt http://astronomy.nmsu.edu/CLOUDS/}.
In the following we present our observations (Section\,\ref{obs}), 
and their reduction and how we search for variability (Section\,\ref{red}). 
In Section\,\ref{anal} we detail for each target the variability information we obtained from our time series, and present variability upper-limits. 
Finally in Section\,\ref{discussion} we interpret our results and link them to atmospheric physical processes.


\section{Observations} \label{obs}

\subsection{Target selection}

We chose our targets according to several criteria.
The targets had to sample the L/T transition spectral types, from late L to early T. 
The $J$~band magnitude of the target had to be brighter 
than 15.5, in order to obtain a high signal-to-noise ratio (SNR) over most of the wavelength range, with a good time sampling.
{SDSS\,J1254$-$0122} was a natural target as we had detected photometric variations in $J$, $H$ and MKO~$K$ bands in April~2002 (Goldman et~al.~\cite{Gol03}).

We selected brown dwarfs having a nearby star within {90\,\arcsec}.
VLT/{\,\sc Isaac} has a long slit of {2\arcmin} that allowed us to simultaneously observe a second star (hereafter: comparison star). 
This simplifies and strengthens the analysis as the comparison star can be used to remove most of the sky absorption variability that affects the target spectrum.
We required the comparison star to be brighter than the target in the $J$~band.
The actual ratios of the comparison star flux to the target flux  at 1.08\,$\mu$m vary between 2 ({SDSS\,J1254$-$0122}) and over 30 ({2MASS\,J0825+2115}).
These constraints limit the number of possible targets to just a few.

A T6 brown dwarf, {SDSS\,J1624+0029}, was selected {to compare L/T transition object variability characteristics with later-type objects,}
and because it was reported variable based on near-infrared spectroscopic observations (Nakajima et~al.~\cite{Nak00}).
We also observed the mid-T dwarfs {2MASS\,J1534$-$2952AB} (T5) and {2MASS\,J1225$-$2739AB} (T6),
when observing conditions prevented us from monitoring other targets.  These two binaries are not resolved in our observations.

Table~\ref{tab_targets} presents the final list of targets with a short description, including binarity and previous variability information.
In this article, we use {near-infrared classification schemes of Burgasser et~al.~(\cite{Bur06}) and of Geballe et~al.~(\cite{Geb02}) in the case of {2MASS\,J0825+2115}.}
The target names are truncated to the first four digits of the right ascension and declination.

\begin{table*}
  \caption[]{Summary of targets and references. } 
  \label{tab_targets}
  \begin{tabular}{lclccccc}
    \hline
    \hline
    \noalign{\smallskip}
    Target &                          {\sc SpeX}   & Sp.T.$^1$ & $J$       & $M_J$             & separation$^2$  & discovery & variability \\
    \noalign{\smallskip}
    \hline
    \noalign{\smallskip}
    2MASS\,J08251968+2115521      & yes & L6           & 14.89     & $14.84\pm 0.08^3$ &     ---$^4$ & Kirkpatrick et~al.~(\cite{Kir00})      & \\ 
    2MASS\,J12255432$-$2739466A   & \ldots & T6           & 15.50    & $14.92\pm 0.14^3$ & $282\pm 5$\,mas & Burgasser et~al.~(\cite{Bur99}) & \\ 
    2MASS\,J12255432$-$2739466B   & \ldots & T8           &  16.85   & $16.27\pm 0.14^3$ & \ldots & Burgasser et~al.~(\cite{Bur03}) & \\
    SDSS\,J125453.90$-$012247.4   & yes & T2           & 14.66     & $14.00\pm 0.06^5$       &   --- $^6$          & Leggett et~al.~(\cite{Leg00}) & J, H, K$^8$\\ 
    2MASS\,J15344984$-$2952274AB  & \ldots & T5.5 & $14.60^5$ & $13.94\pm 0.06^7$ & $65\pm 7$\,mas & Chiu et~al.~(\cite{Chi06}) & \\ 
    SDSS J162414.37+002915.6            & \ldots & T6           & 15.41$^3$ & $15.13^{+0.06}_{-0.05}\,^3$ & \ldots & Strauss et~al.~(\cite{Str99})& H$_2$O$^9$\\ 
    \noalign{\smallskip}
    \hline
  \end{tabular}
  \begin{tabular}{l}
  All targets were observed with {\sc Isaac}. \\
  $^1$ All spectral types from Burgasser et~al.~(\cite{Bur06}) except for {2MASS\,J0825+2115} (Geballe et~al.~\cite{Geb02}).  \\ 
  \hspace{1.3mm} {The photometric type of {2MASS\,J12255432$-$2739466B} was determined from photometry, and is hence uncertain.} \\
  $^2$ Binary separations, 2MASS~$J$ magnitudes and spectral types for each component from Burgasser et~al.~(\cite{Bur03}). \\ 
  $^3$ 2MASS $J$ photometry from Vrba et~al.~(\cite{Vrb04}). \\
  $^4$ Unresolved (Bouy et~al.~\cite{Bou03}).\\ 
  $^5$ MKO $J$ magnitude from Knapp et~al.~(\cite{Kna04}). \\
  $^6$ Unresolved (Burgasser et~al.~\cite{Bur06b}, Goldman et~al., {\it in prep}).\\
  $^7$ MKO $J$ magnitude from Tinney et~al.~(\cite{Tin03}). \\
  $^8$ Goldman et~al.~(\cite{Gol03}) reported photometric variations in $J$, $H$ and $K$ bands. See Section~\ref{sd1254}. \\
  $^9$ Nakajima et~al.~(\cite{Nak00}) reported spectroscopic variations in the water bands over 1.53--1.58\,$  \mu$m. \\
  \end{tabular}
\end{table*}

\subsection{VLT/{I\small SAAC} spectra}

We used {\sc Isaac}, the Infrared Spectrometer And Array Camera mounted on {\sc VLT--Antu}, during the nights of 2003, April 16 and~17.
Table\,\ref{ISAACobs} shows the observing log.
The low-resolution grism covers the wavelength range of \mbox{0.99--1.13\,$\mu$m} 
with a resolving power of $R\equiv \frac{\lambda}{\Delta\lambda} \approx 550$.
Both nights were partly cloudy. 
Relative extinction {and slit losses} for each target were up to 0.55 and 1.0\,mag for
April 16 and 17, respectively.
When we observed the Northern targets 
({2MASS\,J0825+2115}, {SDSS\,J1254$-$0122}, {SDSS\,J1624+0029}),
the winds, from the North, made the seeing highly variable.
We observed {2MASS\,J1534$-$2952AB} and {2MASS\,J1225$-$2739AB} 
when stronger winds prevented us from observing towards the North.
Most observations were conducted using the 1\arcsec-wide slit {(corresponding to a 7-pixel resolution element)}, 
except 
at the beginning and the end of the second night when we used the 1\farcs5-wide slit {(10-pixel)}.
The site seeing during the first night varied between 0\farcs6 to~1\farcs2,
increasing sharply at the end of the {SDSS\,J1254$-$0122} observations and the beginning of the {2MASS\,J1225$-$2739AB} observations.
During the second night, 
the seeing increased slowly from 0\farcs95 to {1\farcs2}.

Spectra were chopped 
between two positions, A and B, separated by 20\arcsec.
In order to minimize the impact of bad pixels and the systematic effects of inaccurate flat fielding,
the A and B positions were randomly moved around each center by up to 5\arcsec.
We chose a single integration time of 2.5\,min to minimize sky brightness fluctuations between the A and B exposures.
All targets were observed with a comparison star in the slit, 
and this constraint sets the slit angle. 
The parallactic and slit angles are indicated in Table\,\ref{ISAACobs}.
The centering of the stars was typically checked after two~hours of observations. The centering procedure causes the grism wheel to be moved; no other changes in instrument configuration occurred during each observing night.

\begin{figure}[th]
\centering
\includegraphics[width=.5\textwidth]{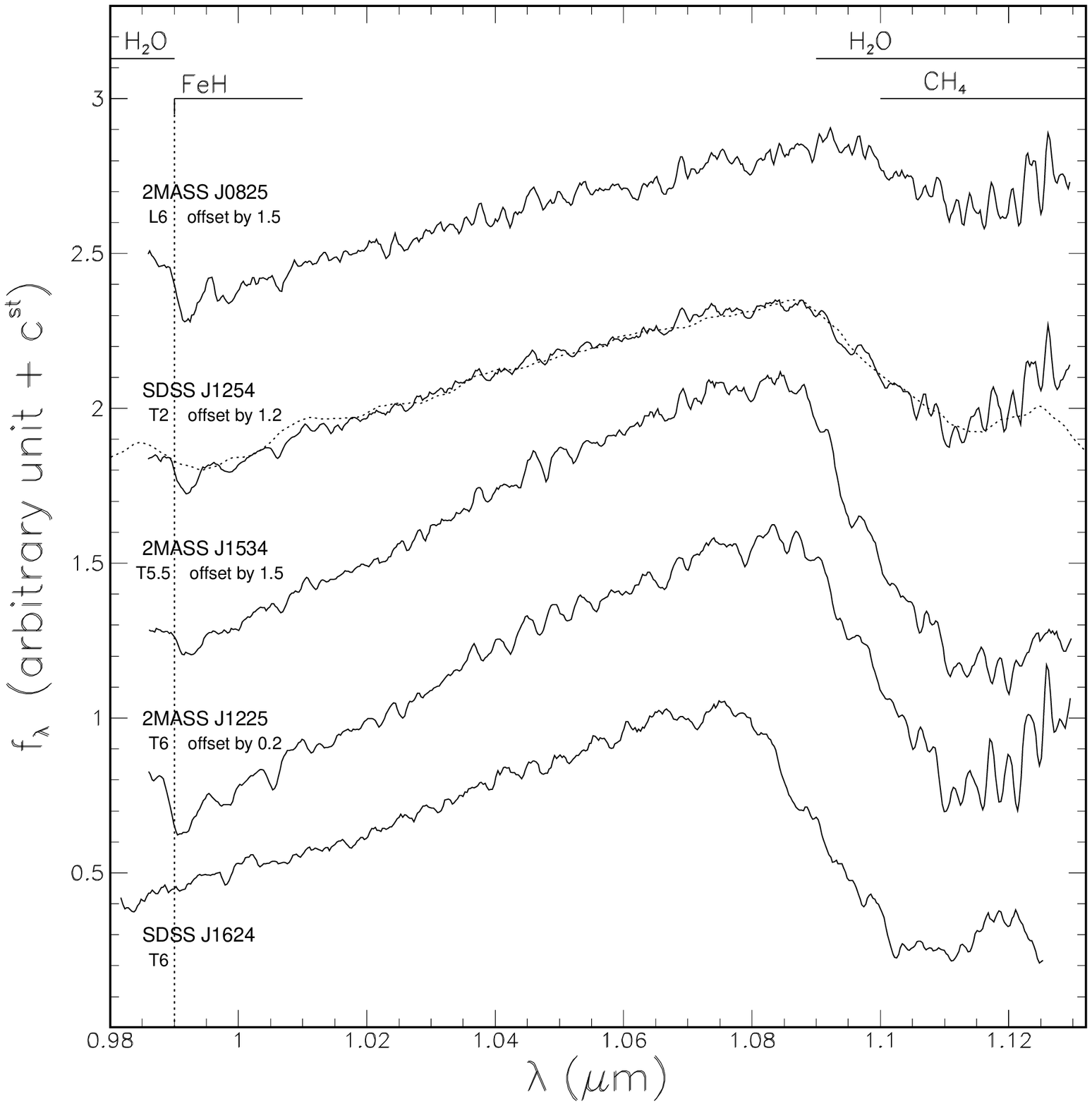}
\includegraphics[width=.5\textwidth]{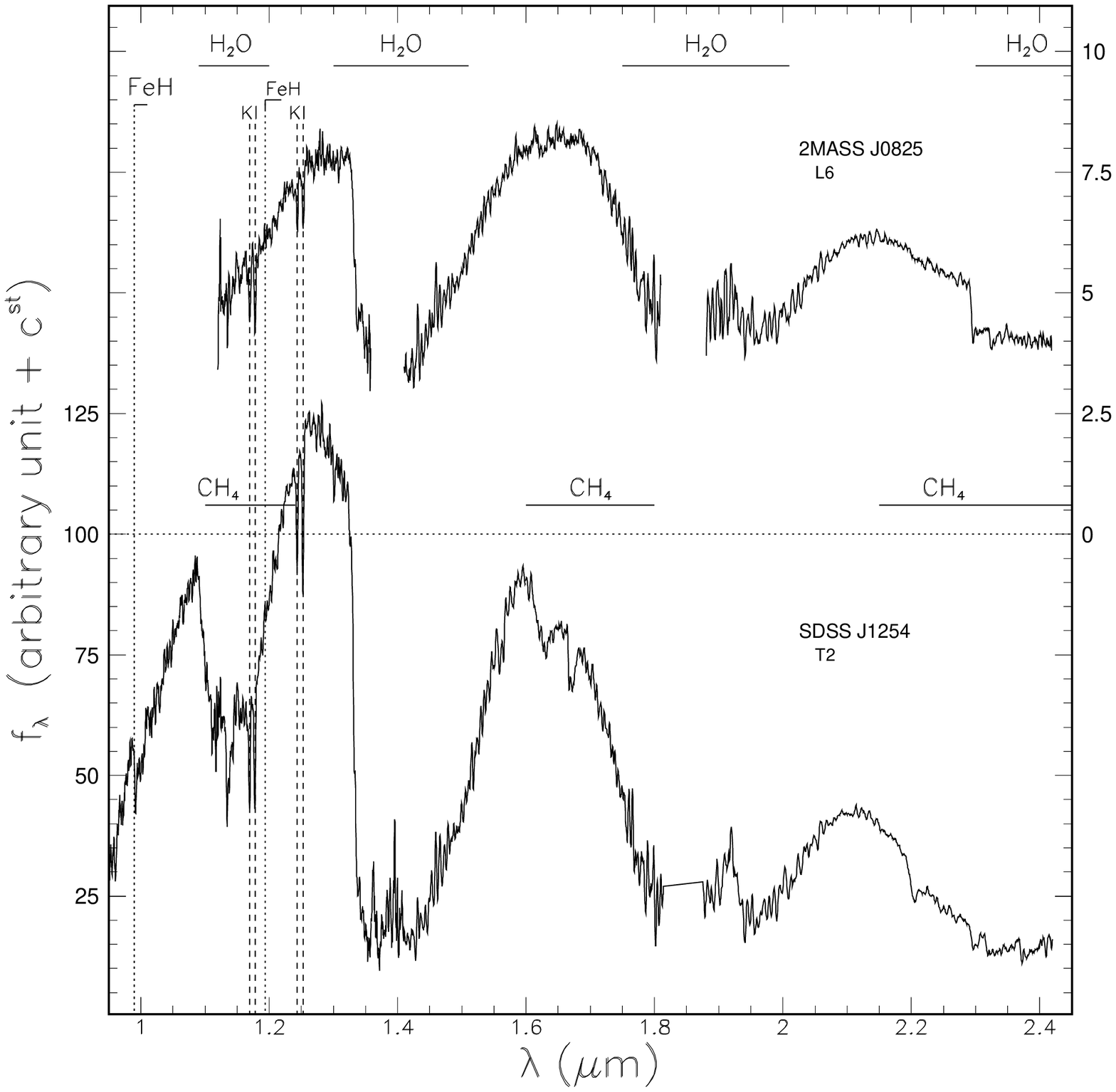}
   \caption{{\em Top:} {\sc Isaac} spectra of April 16 (April 17 for {SDSS\,J1624+0029}) for our targets, 
            ordered by spectra type. The spectra are wavelength and flux calibrated {using an A0 star},
            smoothed over 1\arcsec.
            The spectra are offset for clarity, as indicated in the figure.
            The smoothed IRTF {SDSS\,J1254$-$0122} stacked spectrum is superimposed.
            {\em Bottom:} {\sc SpeX} spectra averaged over all nights for our two IRTF targets.
            Note the different vertical scales for {2MASS\,J0825+2115} (right labels) and {SDSS\,J1254$-$0122} (left labels). 
           We identify the most significant absorption features.
           }
      \label{spectraVLT}
      \label{spectraIRTF}
\end{figure}

After the last scientific observations of each target,
except for {2MASS\,J0825+2115} at the beginning of the second night
and {2MASS\,J1534$-$2952AB} at the end of the second night,
we took a lamp flat, a bias, 
and a series of arcs for wavelength calibration.
A spectrometric standard, the A0~V star HD\,11174, was occasionally observed 
(once the first night, twice on the second night). 
As we are not interested in absolute spectra, we only took these for future reference.
The stacked VLT spectra of April 16 (April 17 for {SDSS\,J1624+0029}),
wavelength- and flux-calibrated, are shown in Fig.\,\ref{spectraVLT} (top).
{The SNR of the spectra over 1.0--1.1\,$\mu$m ranges from 100 to 200
(75 for {SDSS\,J1624+0029}).
The features seen over 1 to 1.1\,$\mu$m are mostly real.
Several features in the {SDSS\,J1254$-$0122} spectrum match those of the NIRSPEC spectrum published by McLean et~al.~(\cite{McL03}). 
Others are barely significant and may be due to systematic errors or noise.
}

\begin{table*}
  \caption[]{{\sc Isaac} observing log. The reported seeing values are those of the DIMM seeing monitor; seeing measured on the frames show very similar values.}
  \label{ISAACobs}
  \begin{tabular}{llcccrc@{ }cc}
    \hline
    \hline
    \noalign{\smallskip}
             &                     &           & slit      &                      &          & \# of spectra & slit               & parallactic         \\
    Date     & Target              &  Start UT & width     & site seeing          & Duration &  [\# taken]   & angle (\degr) & angle (\degr) \\
    \noalign{\smallskip}
    \hline
    \noalign{\smallskip}
    April 16 & {2MASS\,J0825+2115}   & 00:10:49 & 1\arcsec   & 0\farcs82--0\farcs55 & 70\,min  &  20           & $114$            & 167--148\\
             & HD\,11174            & 01:30:20 & 1\arcsec   & 0\farcs55            & 1\,min   &   2           & $179$               & 144--49\\
             & {SDSS\,J1254$-$0122}  & 01:42:05 & 1\arcsec   & 0\farcs80--1\farcs05 & 208\,min &  66           & $141$              & 45--145 \\
             & {2MASS\,J1225$-$2739AB} & 05:30:33 & 1\arcsec   & 0\farcs80--1\farcs25 & 90\,min  &  30           & $54$             & 90--100\\
             & {2MASS\,J1534$-$2952AB} & 07:49:22 & 1\arcsec & 0\farcs95--1\farcs40 & 144\,min &  50           & 98                 & 76--97 \\
    \hline
    \noalign{\smallskip}
    April 17 & {2MASS\,J0825+2115}   & 00:17:01 & 1.5\arcsec & 0\farcs80--1\farcs10 & 56\,min  &  20           & $114 $            & 164--149 \\
             & HD\,11174           & 01:23:52  & 1.5\arcsec & 0\farcs90--1\farcs10 & 15\,min  &   8           & $179$               & 44--41\\
             & {SDSS\,J1254$-$0122} & 01:50:17 & 1.5\arcsec & 0\farcs80--1\farcs18 & 103\,min &  36           &  $141$             & 50--12 \\
             & {SDSS\,J1254$-$0122} & 03:49:06  & 1\arcsec   & 0\farcs85--1\farcs22 (1\farcs40) & 163\,min & 44 [52] & $141$             & 182--124\\
             & HD\,11174           & 06:40:53  & 1\arcsec   & 1\farcs02--1\farcs24 & 2\,min   &   4           & $179$               & 131\\
             & {SDSS\,J1624+0029}    & 07:26:58  & 1.5\arcsec & 0\farcs72--1\farcs40 & 150\,min &  30 [42]      & 30                 & 178--127 \\
    \noalign{\smallskip}
    \hline
  \end{tabular}
\end{table*}

\subsection{IRTF/S{\small pe}X spectra}

We obtained spectra of {2MASS\,J0825$+$2115} and {SDSS\,J1254$-$0122} using {\sc SpeX} (Rayner et~al.~\cite{Ray03}), 
the medium-resolution cross-dispersed spectrograph mounted on the 3.0\,m NASA Infrared Telescope Facility (IRTF), on 2003 April 18 to 21 (UT). 
{Table\,\ref{SpeXobs} shows the observing log.}
The stacked IRTF spectra of April 18 and 19 for {2MASS\,J0825+2115} 
(April 18, 20 and 21 for {SDSS\,J1254+0122}),
wavelength- and flux-calibrated, are shown in Fig.\,\ref{spectraIRTF} (bottom).
Observing conditions were non-photometric on all nights and observations were often halted due to thick cloud coverage. 
The seeing was typically {$\sim$0\farcs8} in the $H$-band. 
We used the ShortXD mode with the 0$\farcs$8-wide slit that covers from 0.8$-$2.4 $\mu$m at \mbox{$R\approx 750$} in a single exposure. 
All observations were conducted at the parallactic angle to minimize differential slit losses.
{No comparison stars were observed simultaneously because of the 15\,\arcsec slit length.}

For each target, we obtained a series of 2-min exposures. 
During the series, the telescope was nodded between two positions along the 15$\arcsec$ slit to facilitate the subtraction of the dark current and sky background in the data reduction process. 
Observations of an A0~V standard star were also acquired to correct each target for telluric absorption. 
The airmass difference between the targets and the A0~V stars was always less than 0.1. 
Finally, a set of calibration frames including flat field and argon arc exposures were obtained for each target--standard star pair. 

\begin{table}
  \caption[]{{\sc SpeX} observing log. All slit angles were parallactic. Slit width was 0\farcs8. Spectroscopic standard observations, between sets of target observations, are not reported. }
  \label{SpeXobs}
  \begin{tabular}{llcrc@{ }}
    \hline
    \hline
    \noalign{\smallskip}
    Date     & Target               &  Start UT  & Duration & spectra  \\
    \noalign{\smallskip}
    \hline
    \noalign{\smallskip}
    April 17 & {2MASS\,J0825+2115}   & 06:19:38   & 42\,min  &  18         \\
    \hline
    April 18 & {2MASS\,J0825+2115}   & 05:35:48   & 18\,min  &  8         \\
             & {SDSS\,J1254$-$0122}   & 08:32:05   & 42\,min &  18     \\
             &                              & 09:31:41   & 42\,min &  18     \\
             &                              & 10:25:49   & 28\,min &  9     \\
             &                              & 12:30:03   & 18\,min &  8     \\
    \hline
    April 19 & {2MASS\,J0825+2115}   & 05:33:27   & 20\,min  &  10         \\
             &                              & 06:15:03   & 21\,min  &  9         \\
             &                              & 07:43:54   & 42\,min  &  18         \\
    \hline
   April 20  & {SDSS\,J1254$-$0122}   & 10:07:23   & 21\,min &  10     \\
             &                              & 10:45:46   & 42\,min &  20     \\
             &                              & 10:50:29   & 26\,min &  11     \\
             &                              & 12:27:47   & 22\,min &  10     \\
    \hline
   April 21  & {SDSS\,J1254$-$0122}   & 10:43:02   & 20\,min &  10     \\
             &                              & 11:21:41   & 41\,min &  20     \\
             &                              & 12:19:58   & 22\,min &  10     \\
    \noalign{\smallskip}
    \hline
  \end{tabular}
\end{table}

\section{Reduction} \label{red}

\subsection{VLT/{\,\sc Isaac} spectra}

We reduce the spectra following the {\sc Isaac} Data Reduction Guide\,1.5.
We remove the electric ghosts generated by the HAWAII detector using the ESO {\sc eclipse ghost} recipe. 
The next steps are performed using a modified version of reduction package 
{\sc Spextool} v3.1 (Cushing et~al.~\cite{Cus04}). 
We subtract the A and B images to remove the sky background and bias, and the resulting images are flat-fielded. 
The creation of the flat field is a crucial step and its accuracy will be studied in Section~\ref{ISAACanal}.
The residual sky background is subtracted, 
and the spectra are extracted, using an aperture independent of wavelength. 
{For each target, 
we wavelength calibrate each spectrum relative to a reference sky spectrum, 
before we calculate the wavelength solution for this reference spectrum.
We extract a sky spectrum corresponding to the A position from the B image,
and shift the A object spectrum by a constant, using the OH~lines of the sky spectra.
{Because this constant is usually not an integer, in units of wavelength-pixels, the flux values are linearly interpolated for the resulting fractional-pixel shifts. 
We estimate the errors introduced by this process to be about 1\,\% in rapidly changing parts of the spectrum (water bands) and much less elsewhere.}
}
For the two targets observed both nights, a shift of about 15\,pixels occurred between the two nights of observation, much larger than shifts within a night.

The reference sky spectra are calibrated to actual wavelengths using the arc spectra taken right after the observations, and checked against OH sky lines, with a 3--4\,{\AA} agreement.
Absolute calibration is only required for line identification, not variability detection.

The target spectra are then divided by the comparison star spectrum.
Because we observed over 0.99--1.13\,$\mu$m where the Earth atmospheric transmission is good,
the main effect of this correction is to remove terrestrial grey atmospheric absorption (clouds),
instrumental response, 
and {slit losses, mostly due to seeing variations}.
The comparison star, however, is useful to remove the water-band variations redward of 1.10\,$\mu$m.
The resulting spectra are co-added by groups of six to~eight (15 to 20\,minutes),
depending on the target.
It proved unnecessary to apply a median filter or clip the data, 
as most of the cosmic rays and bad pixels are removed by {\sc Spextool} during the spectrum extraction.
We choose to co-add six to~eight spectra into stacked spectra, in order to get sufficient SNR, 
while maintaining sensitivity to short term variations. 
We search for variability on these 15- to 20-min spectra,
divided by the rest of the night data set.
Finally, the overall flux level (independent of wavelength) may vary differently for the target and the comparison star, whenever the centering of both objects is not identical, so that the amount of flux losses differs.
Depending on targets (and observing conditions), we sometimes have to normalise each spectrum over a broad wavelength region, {when the stability obtained after correcting with the comparison star's spectrum  is not good enough for our purposes.  
For homogeneity, we apply this normalisation to all VLT targets.
To determine that region, we try to maximise its extend, in order to be sensitive to variations affecting large parts of our wavelength coverage, and to increase the accuracy, while avoiding zones of artefacts (ghosts) and low signal (water bands, order edges).
The two latter requirements vary with targets, and the exact values are given in the legend of Fig.\,\ref{fig-2m0825vlt},\,\ref{fig-sd1254vlt},\,\ref{fig-2m1534},\,\ref{fig-2m1225} and \ref{fig-sd1624}.
The resulting regions are 38-nm\ wide (27\,\% or our wavelength coverage, for {SDSS\,J1624+0029}) or wider. }
Searching for broad band photometric variations (outside atmospheric water bands) is in any case better performed with imaging.
The dispersion of the stacked spectrum flux values is used as a measure of the (total) error (see \ref{errors}).
The spectra presented in 
Fig.\,\ref{fig-2m0825vlt},\,\ref{fig-sd1254vlt},\,\ref{fig-2m1534},\,\ref{fig-2m1225} and \ref{fig-sd1624}
are smoothed over 11\,pixels, or 1.5 times the actual resolution obtained with the 1\arcsec-wide slit. 

\subsection{IRTF/{\,\sc SpeX} spectra}

The data are reduced using {\sc Spextool} (Cushing et~al.~\cite{Cus04}), the IDL-based data reduction package for {\sc SpeX}.  
Pairs of exposures taken at the two different slit positions are first differenced to remove the dark current and sky background (to first order) and then flat fielded.  
The spectra are extracted, using an aperture independent of wavelength, and wavelength calibrated after subtracting any residual sky background. 
The raw spectra are then corrected for telluric absorption using the A0~V star spectra and the technique described by Vacca et~al.~(\cite{Vac03}).  
Finally, the telluric-corrected spectra from the different orders are merged to create continuous $\sim$1$-$2.4 $\mu$m spectra for each target. 

As for the VLT spectra, the dispersion of the flux values is used as a measure of the error {(see \ref{errors} for a brief discussion of the drawbacks of this method)}.
We also normalise each spectrum over J {(more exactly 1.19--1.32$\,\mu$m), H (1.44--1.75$\,\mu$m) and K (2.02--2.19$\,\mu$m)}, and perform our variability search independently in each  region of the spectrum.

The spectra presented in Fig.\,\ref{fig-2m0825irtf} and \ref{fig-sd1254irtf} are smoothed over 13\,pixels.

\subsection{Slit losses, refraction and the like}

{Varying observing and instrumental conditions (such as movements of the slit relatively to the target, or seeing variations) occurring during the observations could be a problem for our study.
Because we aim at a great precision (better than 1\,\%) of our relative spectroscopy, 
we need to pay special attention to flux losses that could mimic intrinsic variations.
{The effects described in this section vary smoothly with wavelength. They will not affect the analysis of high-frequency variations, but mostly the calculations based on the spectral indices of Section\,\ref{model}.}

Indeed, the slit cuts the wings of the point-spread function in variable amounts depending on the wavelength.
Seeing varies with wavelength, according to: $s\propto \lambda^{-0.2}$.
Over the {\sc Isaac} wavelength range, this means that the flux loss is smaller by 1\,\% at $1.13\,\mu$m relative to $0.99\,\mu$m (assuming 1\arcsec\ seeing and 1\arcsec\ slit width, and the stars to be centered on the slit). 
This effect, however, affects both the target and the reference star equally and is therefore removed.
Within the {\sc SpeX} $J$, $H$ and $K_s$ bands, the effect is 2\,\% to 1\,\%, for 0.8\arcsec\ to 1.2\arcsec\ seeing (0.8\arcsec\ slit width). Most of this is removed using the reference stars, which were observed every 20 to 40\,min, hence under similar seeing conditions. 

The stellar centroid also shifts with wavelength due to atmospheric refraction, depending on the hour angle.
For the IRTF data, we observed at the parallactic angle, so that the star was equally centered on the slit at all wavelengths. The main limitation here is the wavelength-dependent variations of the atmospheric transparency between two observations of the reference star.

For the VLT data, the slit angle was set by the nearby reference star (see Table\,\ref{ISAACobs}). We are now concerned with differential, wavelength-dependent flux losses between the target and the reference star.
Slit translation movements and seeing variations, if the target and the reference star are not equally centred in the slit, and slit rotation, will result in wavelength-dependent flux losses, due the wavelength-dependent centroids.
(If the two objects are equally centred on the slit, flux losses will be similar and cancel out.)
Slit rotations are not expected with {\sc Isaac}. 
Occasional flexures are not known to produce significant effect with our slit widths. 
For two targets, we performed a second acquisition to check orientation after a long observing sequence, and found no significant rotation.
The VLT unit telescopes are altazimutal but none of our targets were observed closer than $17\deg$ to the zenith, while $10\deg$ is considered a safe limit. 

We nevertheless investigate the possibility of a colour effect of a small rotation.
We estimate the differential refraction shift using the procedure of Marchetti\footnote{\tt eso.org/gen-fac/pubs/astclim/lasilla/diffrefr.html}. 
The maximal relative shift predicted between $0.99\,\mu$m and $1.12\,\mu$m is 0.05\,\arcsec, 
when comparing observations taken at the minimal and maximal airmasses of each target.

In practice, our spectral index $s_{\rm VLT}$ 
(see \ref{model}) does not show any correlation with the normalisation factor, for any target.
We conclude that refraction and slit losses do not add systematic effects to our measurement. 
They may increase the statistical noise, although this will only be {perceptible} as we add fluxes of many pixels.
}

\section{Analysis} \label{anal}

\subsection{General considerations on the VLT/{\,\sc Isaac} spectra} \label{ISAACanal}

The correction by the comparison star allows us to obtain a 1--3\,\% precision in most co-added spectra, 
before normalisation. 
{This good result proves the usefulness of observing a comparison star simultaneously with the target.}
The {\sc Isaac} flat fields are known to be very stable over several nights for high frequency scales.
The set up (grism, etc) wasn't changed during the observations.
Only the grism wheel was rotated to center the star in the slit about every second hour.
The instrument is on a Nasmyth focus which reduces the distortions, 
{such as wavelength shifts.}
Flat field errors are an important source of systematic errors.
{We compare the flat fields taken during the same night, with a few hour gap. 
A slope of about 0.5\,\% and broad variations of similar amplitude are observed.
We do not remove these effects.}
 They are always insignificant compared to the photon noise of the brown dwarfs.
When we stack the spectra of the brightest comparison stars, 
we find that the flux variations over 0.99 to~1.11\,$\mu$m are
0.43\,\%, 0.52\,\% and 0.45\,\% (per 
pixel) for the comparison stars of {2MASS\,J0825+2115}, {2MASS\,J1225$-$2739AB} and {2MASS\,J1534$-$2952AB}, respectively, with no slope larger than 0.05\,\% per 0.1\,$\mu$m.

Flat fields taken the same night, but for different targets or different slit widths,
do show a 5\,\% variation at 1.09\,$\mu$m. 
As this is seen over the whole chip, 
this effect is mostly removed by dividing the spectra by a median spectrum.
Because we generally used flat fields taken right after the observations,
it is possible that this effect does not affect our data.
{Consequently, systematic effects are likely to be} less than, or occasionally equal, to 0.5\,\%.

The flat fields between two nights vary much more, due to day-time instrument configuration changes for routine calibration (grism position, slit, flat field lamp temperature). 
Although some broad-band effects can be removed by fitting a 3--5-degree polynomial, 
the systematics increase significantly, {and we did not correct for these variations.} 

A ``ghost'' of each spectrum, scaled by about 1.5\,\%, is present at $\sim$$1024-y$ pixels where $y$ is the position of the spectrum along the spatial axis, over 525--635\,{rows} in the dispersion axis (about 1.055--1.086\,$\mu$m). This 
 is probably an internal, optical reflection, or an electric cross-talk.
The comparison star's ghost overlaps with a spectrum (comparison star or target) for about 19~spectra, which are only used over the unaffected region of the spectrum.
About 20~ghosts overlap with a spectrum of the corresponding dithered frame.
We examined those spectra and masked the affected wavelengths if the distance between the ghost of the brightest comparison star and the target spectrum is smaller than 2\arcsec.
We could have subtracted a different image, but that would have increased the time difference between the pair of images and the contribution to the noise of the sky subtraction.
The effect of the target ghost on the faintest comparison star is 0.7\,\% at most and is neglected.

\subsection{Flux uncertainties} \label{errors}

In order to estimate the significance of possible variations, we need to know the noise associated with our measurements. 
We use the dispersion observed over 15 to 20\,minutes ({\sc Isaac}) and over 16 to 40\,minutes ({\sc SpeX}). 
This dispersion may include short-term intrinsic variability. 
{It also includes any remaining systematics affecting these short time scales.}
We would not be sensitive to~1\,\% variations over {a few minutes}, due to the limited SNR, anyway.
Given the small number of spectra from which the dispersions are measured (six to eight for {\sc Isaac}, eight to~20 for {\sc SpeX}),
the uncertainties on the dispersion are large. 
The uncertainties are however reduced when we smooth the spectra in wavelength. 

\subsection{Method} \label{method}

We use coherent methods to detect variations and to calculate variability upper limits, in the case of non detections. 
The methods are coherent because we perform identical variability detections on the observed spectra and, in order to determine the upper limits, on simulated spectra.
We base our search on the $\chi^2$ probability distributions of the time series. 
For each wavelength (for each pixel of a reference spectrum),
we fit the time-dependent fluxes by a constant flux and obtain a $\chi^2$. 
{Using a Savitsky-Golay filter,} the stacked spectra are previously smoothed to match the resolution element (resolving power of~$R=500$). 
We also look for lower frequency {(broader)} variations, and perform a similar search after smoothing the spectra to a resolving power of~$R=100$. 
We decide that a variation is significant when the $\chi^2$ value is compatible with the random distribution hypothesis with a probability lower than 1\,\% (99\,\% of confidence level).
For each target, we comment on those variations in Section\,\ref{results}.

We calculate the integrated flux around those features which appear to vary. 
We obtain a narrow-band photometric time series, and search for {linear} correlation between the fluxes obtained for different features.
We report in Table\,\ref{varlines} those features which show correlated variations, with a probability higher that 99\,\% of confidence level.
{Because we perform these correlation measurements only on a limited set of lines, those that show variations, we expect less than one spurious detection. }
Correlation in features widely separated in the spectrum (hence on the array) strongly argue that the variations are not of instrumental origin, but originate in either the target or the Earth's atmosphere.

To calculate the variability upper limits, we proceed in the manner of Bailer-Jones \& Mundt~(\cite{CBJ01}). 
We simulate mock spectroscopic time series, adding noise according to the observed flux errors.
{For each wavelength, we add to each stacked spectrum} time-dependent variations with increasing amplitudes, until the variations are detected, under the procedure used for the actual data.
We model the time-dependent variations with a sinusoid, positive function: $\delta F = |\sin(\omega t+\phi)|$.
This would be observed (having neglected the limb-darkening effect) in the on-average-favourable case of two anti-podean spots.
Hence we assume that a spot is observable at all times,
with a variable projected area. 
This means that our upper limits are only valid while the target was being observed, not on average over the observing run.
This calculation is more appropriate for our limited time coverage.
An hypothesis of numerous spots, simulated by a sum of randomly-phased sinusoids, would also reduce the expected peak-to-peak amplitude.

{Then}, the smallest amplitude that triggers the variability detection is recorded as the variability upper limit for that wavelength.
The mock variations are time-sinusoid, with random periods (between 1 and 16\,hours) and phases.
We average these maximum amplitudes for various periods and phases.

We perform those calculations for each wavelength independently.
However, real variations would certainly be correlated in wavelength, 
because of the non-zero slit width 
and of the physical and chemical processes in the BD atmosphere causing the variations.
As this latter correlation is largely unknown, our mock variations are not correlated with wavelengths.
{However, we stress that the upper-limits we calculate are fluxes of {\it smoothed} spectra;
hence they refer to variations affecting the whole resolution element (for either of the 100 or 500~resolving powers).
}

{Finally, we introduce in Section\,\ref{model} two spectral indices which maximize our predicted sensitivity to variations caused by changes in the cloud pattern, one for the VLT wavelength region: $s=\frac{1.00-1.02\mu\rm m}{1.06-1.08\mu\rm m}$, one for the IRTF data: $s=\frac{1.46-1.50\mu\rm m}{1.58-1.62\mu\rm m},$ and use them to obtain variability upper-limits.
Although the index definitions were motivated by particular models, the variability upper limits arising from the measurements themselves are of course independent of the models.
}

\section{Results} \label{results}

\begin{figure}[th]
\includegraphics[width=.5\textwidth]{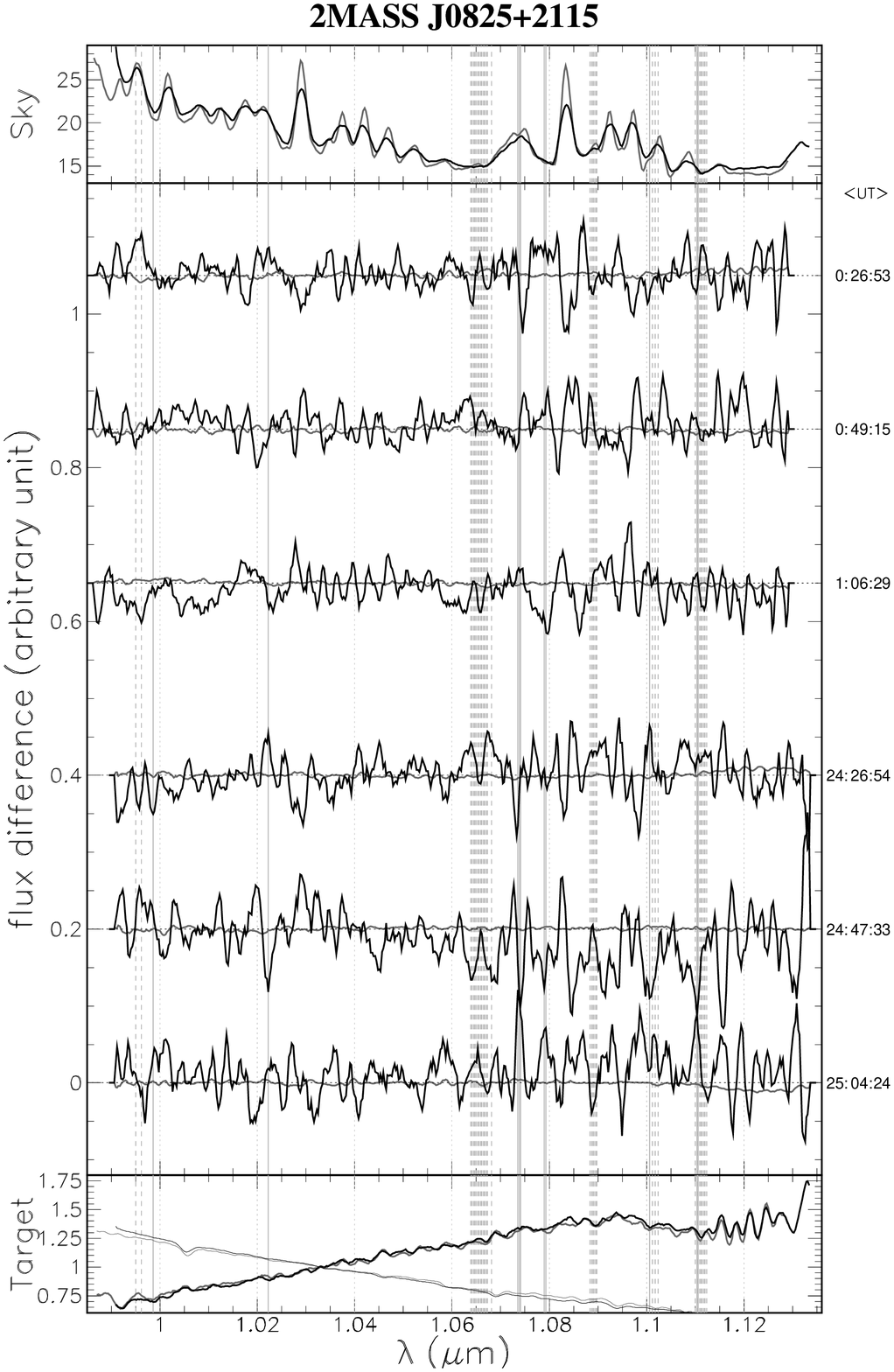}
   \caption{VLT spectroscopic time series for {2MASS\,J0825+2115}. 
            The stacked spectra are relative to the average of each night, 
            and offset for clarity. UT is the mean time since April 16, 2003, UT=0:00.
            The 00:26:53 and 24:26:54 spectra have 20-min exposure time,
            while all other spectra have 15-min exposure time.
            The spectra are normalised over 1.014--1.054\,$\mu$m.
            The brown dwarf flux variations relative to the comparison star are shown as a black (noisy) line.
            The comparison star flux variations are in grey (almost flat) line.
            The variations are relative to the integrated flux of the brown dwarf and the comparison star over that wavelength range.
            The dashed grey lines 
            mark the features that we report to vary 
            at medium resolution while the solid grey lines refer to the lower resolution analysis.
      {\it Top panel:} the sky raw spectra, for both observing nights.
            {\it Bottom panel:} the target raw spectra, for both observing nights.
            The thin-lined bluer spectra are those of the comparison star.
            None are flux-calibrated.
            {\it All panels:} all fluxes are in the {\it same} arbitrary unit, except the comparison star.
           }
      \label{fig-2m0825vlt}
\end{figure}

We now discuss each individual target, starting with earlier spectral types, 
according to the recipes described in Section~\ref{method}.
The spectroscopic time series for each target are presented in Fig.\ref{fig-2m0825vlt} to \ref{fig-sd1624}. 
In this section we report and comment on the detected significant variations.
We discuss the upper limits we obtain in the next section.

\subsection{{2MASS\,J0825+2115} (L6)}

On both April 16 and 17, we obtained a series of twenty 2.5-min {\sc Isaac} spectra (see Fig.\,\ref{fig-2m0825vlt}). 
We combined them by groups of six or eight.
The standard deviations of the 15-min spectra is 1.5\,\% per resolution element with peaks at 3\,\% in the strongest sky lines. 

At the resolution of $R=500$, we find five~features with significant variations, with width of one to three~pixels, all during the second night (see Fig.\ref{fig-2m0825vlt}).
At $R=100$, we find broader and more significant features: at $0.996\,\mu$m during April~16, and more clearly over 1.064--$1.068\,\mu$m, at $1.089\,\mu$m and $1.111\,\mu$m (see Table\,\ref{varlines}).
However, none of those features vary with any significant correlation with any other line.

We also obtained 45 {\sc SpeX} 2-min spectra on April 18 and 19, which we combined by groups of eight to~18, depending on the observing conditions (see Fig.\,\ref{fig-2m0825irtf}). 
The relative precision is about 3--6\,\% in $J$~band, 2--5\,\% in the $H$~band and better than 2\,\% in the $K$~band.
We see one 1-nm wide variable line, at $1.140\,\mu$m. The telluric water band absorption makes there a variability claim difficult, in the absence of simultaneous observations of a comparison star.

The {\sc Isaac} 0.996\,$\mu$m feature could be related to the CrH 0.99685\,$\mu$m bandhead (Kirkpatrick et~al.~\cite{Kir99}) and 0.997\,$\mu$m FeH {Wing-Ford} absorption feature ({Wing et~al.~\cite{Win77}}, Cushing et~al.~\cite{Cus05}).
In cool regions of the photosphere of late-L and T~brown dwarfs, Fe is depleted from the gas to form clouds, that, along with the silicate clouds, have optical depths greater than unity at these wavelengths.
However, when a hole in the cloud deck appears, deeper, warmer regions of the atmosphere are unveiled, so that the FeH absorption features  
might be expected to re-appear and hence vary when the cloud deck evolves (Burgasser et~al.~\cite{Bur02b}).
{Variations in the CrH and FeH bandhead regions redward of 8600\,\AA\ have previously been noticed in the slightly later L8 dwarf Gl~584C over a 3-month baseline (Kirkpatrick et~al.~\cite{Kir01}). However, in the absence of  quantitative analysis of the significance and amplitude of the variations, it is difficult to compare our results.}
The $1.089\,\mu$m and $1.111\,\mu$m features are on the edge of water and methane absorption bands, but no other features appear to vary redward of those features, where absorption is larger, and our sensitivity unchanged (see Fig.~\ref{fig-upperlim-vlt}).

\subsection{SDSS\,J1254$-$0122 (T2)} \label{sd1254}

We obtained a series of 66 and 81 2.5-min {\sc Isaac} spectra, on April 16 and 17 respectively (see Fig.\,\ref{fig-sd1254vlt}).
We combined them by groups of eight, except the last ten of April 16 stacked in one single spectrum. 
The precision ranges from 0.8\,\% to 2\,\% at the strongest sky lines and on the order edges. 

At the resolution of $R=500$, we find a number of features with significant variations, with width of one to four pixels. 
The most significant one is at 0.997\,$\mu$m, mostly on the first night, and less significantly on the second night.
The variations at 1.046\,$\mu$m and 1.11--1.13\,$\mu$m are well detected on both night, while other variations are only observed during one night.
During the second night a total of 16~features appear to vary.
{None of those are strong features with deep unresolved absorption lines, so that telluric differential absorption between the target and the reference star could not explain them.}
At $R=100$, we find no variable features during the first night but four features appear to vary on April~17, particularly at 1.033\,$\mu$m and around 1.104\,$\mu$m (see Table\,\ref{varlines}).
Among those features, we find five pairs which are correlated (see Fig.\ref{fig-sd1254c}, top).

{The most significantly variable features (see Table\,\ref{varlines}) are located close to but outside strong and variable telluric emission lines, such as 0.997\,$\mu$m (close to 0.995\,$\mu$m), 1.031\,$\mu$m (1.029\,$\mu$m), 1.046\,$\mu$m (1.042\,$\mu$m).
Moreover, no variations are detected on the emission lines themselves.
The broadest feature 1.11--1.13\,$\mu$m overlaps with no emission lines.
The observing conditions of {SDSS\,J1254$-$0122} in terms of airmass and atmospheric parameters were similar to, possibly more stable than, those of the rest of the sample\footnote{See \tt http://archive.eso.org/asm/ambient-server
}}.

\begin{figure*}[p]
\centering
\includegraphics[width=.95\textwidth]{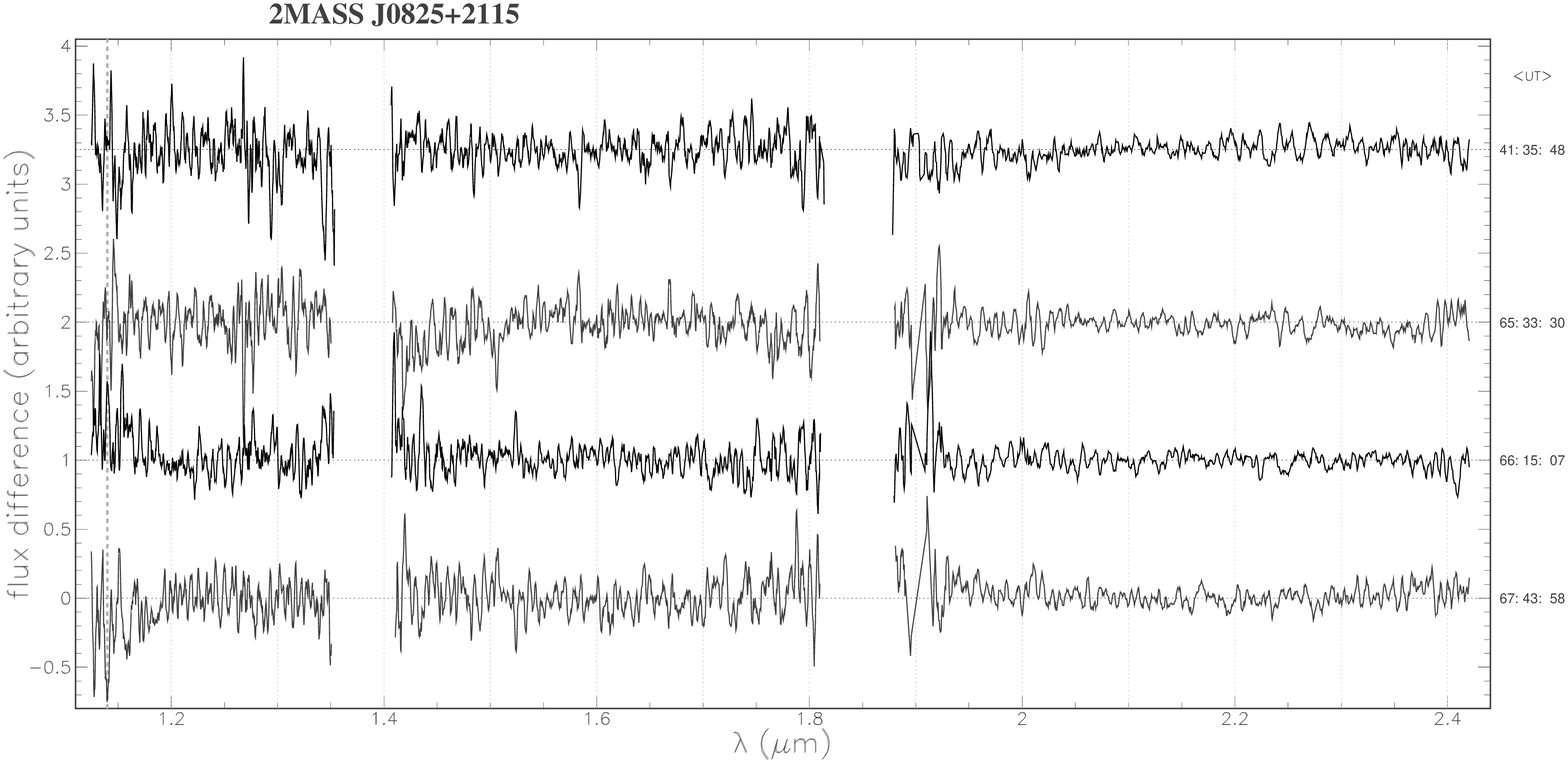}
   \caption{IRTF spectroscopic time series for {2MASS\,J0825+2115}.
            The 16- to 36-min spectra are relative to the average of April 19, 
            and offset. UT is the mean time since April 16, 2003, UT=0:00.
            Measurements with errors larger than $\sim$8\,\%, as well as water bands, 
            are not displayed for clarity.
            No variability information can be confidently obtained in those regions.
            All spectra are normalised independently in $J, H$ and $K$.
            The dashed grey line at 1.140\,$\mu$m marks the feature that we report to vary. 
           }
      \label{fig-2m0825irtf}
      
\includegraphics[width=.95\textwidth]{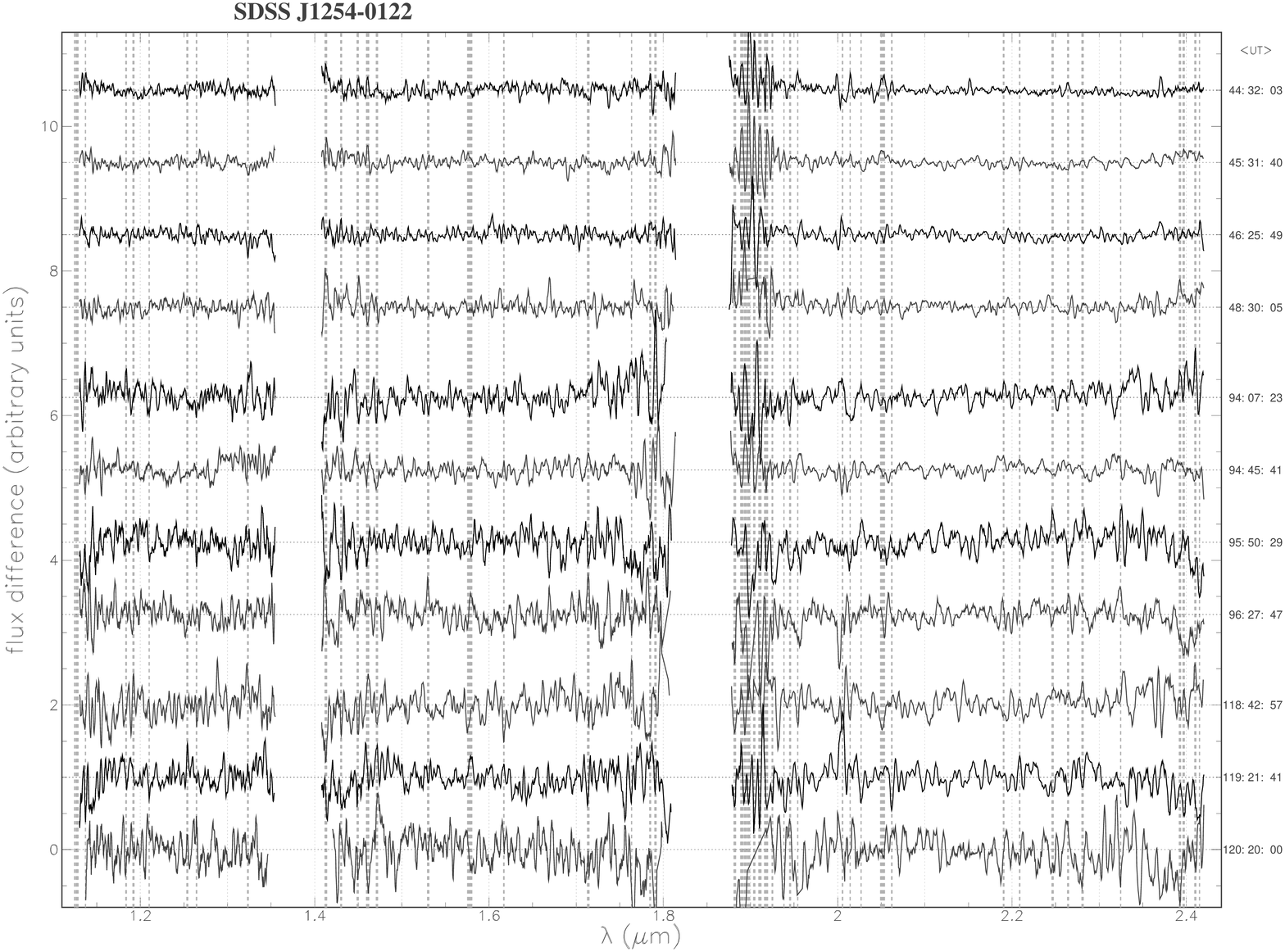}
   \caption{IRTF spectroscopic time series for {SDSS\,J1254$-$0122}.
            The 16- to 40-min spectra are relative to the average of each night, 
            and offset. UT is the mean time since April 16, 2003, UT=0:00.
            See legend of Fig.\,\ref{fig-2m0825irtf} for details.
            {In contrast to {2MASS\,J0825+2115}, many lines exhibit statistically significant variations.}
           }
      \label{fig-sd1254irtf}
\end{figure*}

\begin{figure*}[t]
\includegraphics[width=.5\textwidth]{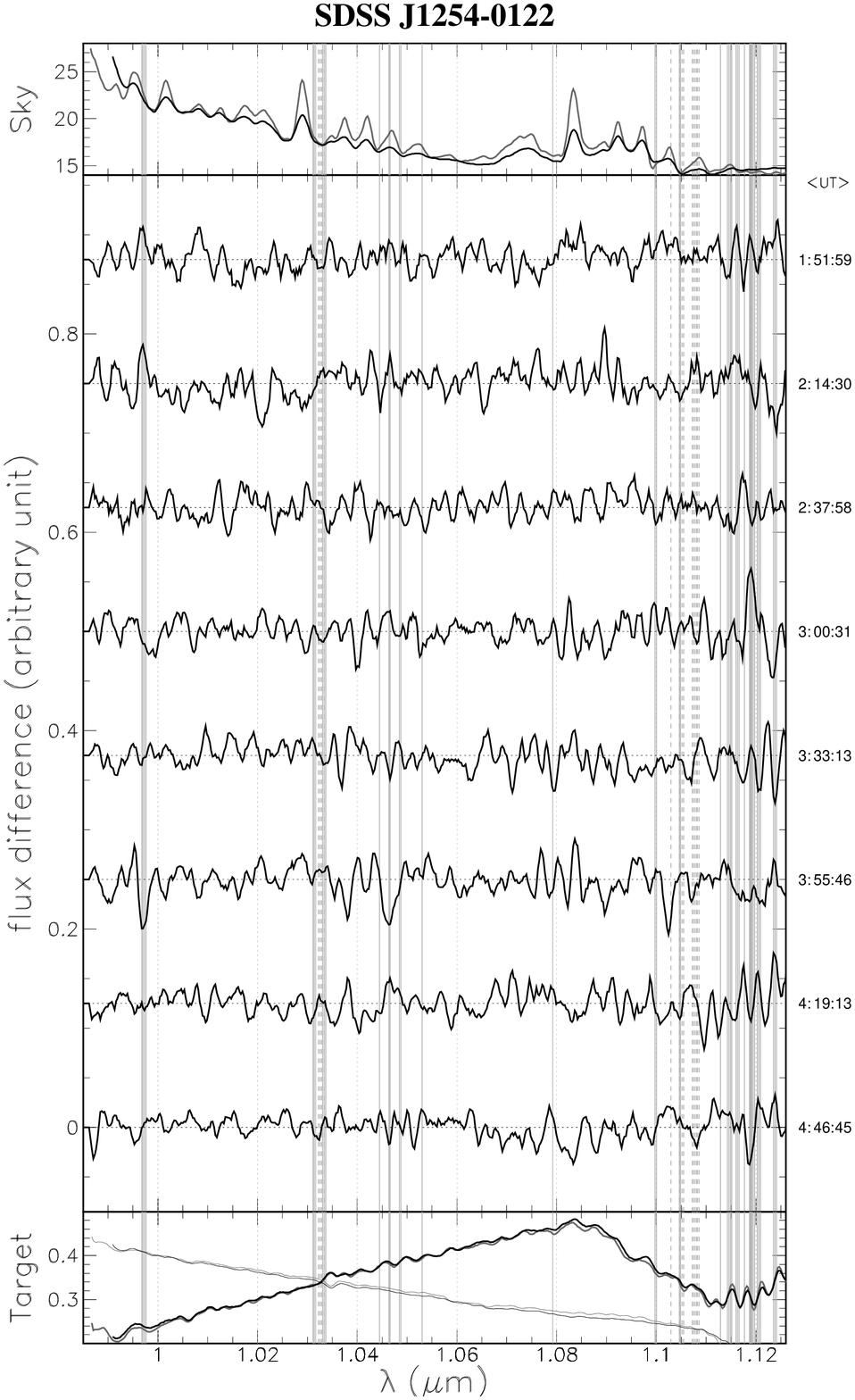}
\includegraphics[width=.5\textwidth]{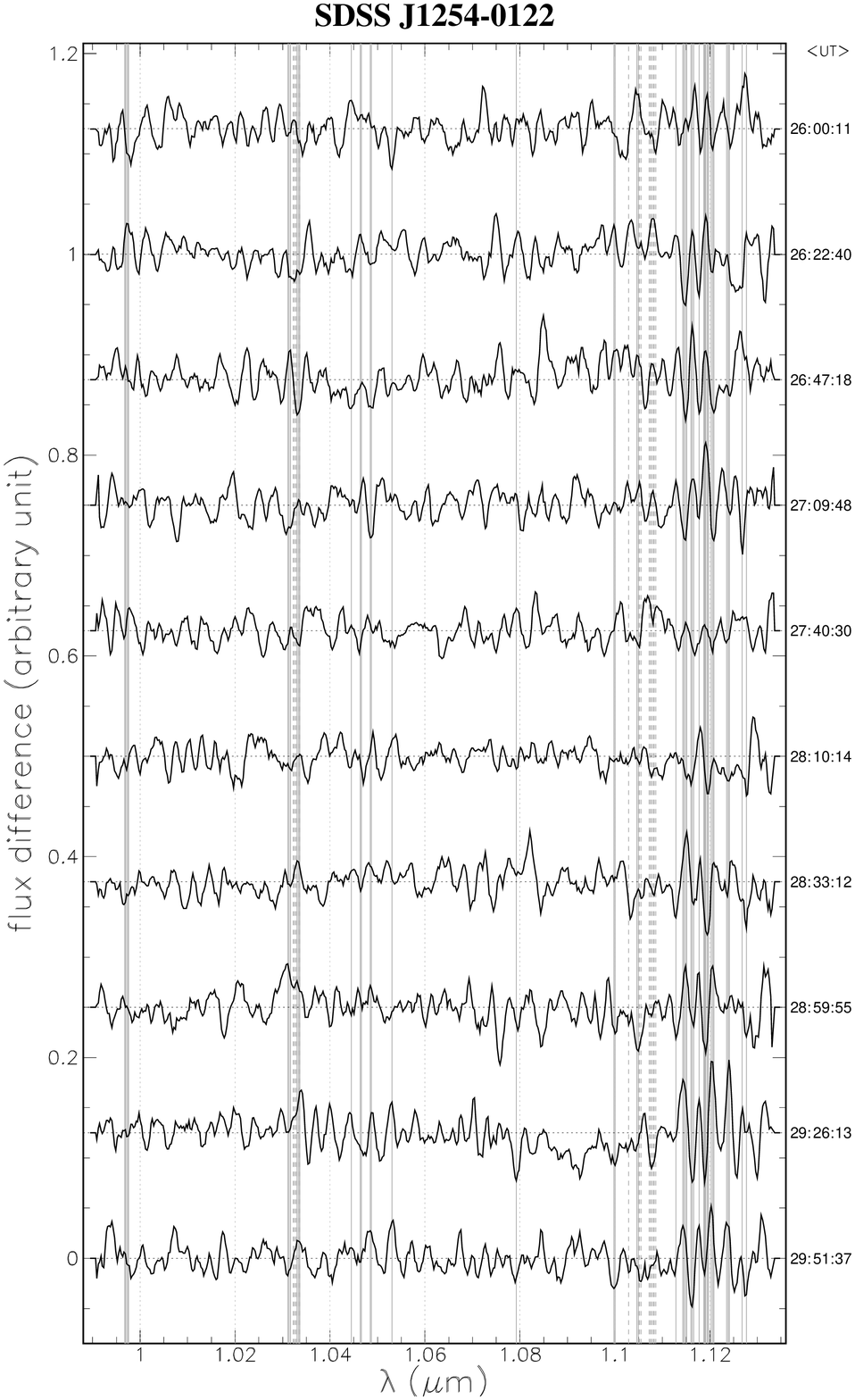}
   \caption{VLT spectroscopic time series for {SDSS\,J1254$-$0122}, for April 16 (left) and 17 (right), 2003.
            The 20-min spectra are relative to the average of each night, 
            and offset for clarity. UT is the mean time since April 16, 2003, UT=0:00.
            The SDSS\,J1254$-$0122  comparison star is only 2.5 to 5 times brighter than the brown dwarf, and its variations are not over-plotted for clarity.
            All spectra are normalised over 1.002--1.088\,$\mu$m.
            The bottom left panel shows the medium signal to noise ratio of SDSS\,J1254$-$0122 as a function of wavelength.
            See legend of Fig.\,\ref{fig-2m0825vlt} for more details. 
           }
      \label{fig-sd1254vlt}
\end{figure*}

We also obtained 146 2-min {\sc SpeX} spectra 
on the nights of April~18, 20 and 21, which we stacked by groups of eight to~20 (see Fig.\,\ref{fig-sd1254irtf}).
At the resolution of $R=500$, we find many features which show variability. 
Among those, a dozen of pairs show significant and convincing correlations.
Correlated pairs cover a wide range of features, with no obvious chemical connections--except possibly water bands, which affect a large part of the near infrared spectrum.
{We see some variations around 1.58$\mu$m in the water bands, as Nakajima et~al.~(\cite{Nak00}) reported in the spectrum of {SDSS\,J1624+0029}, although we find significant variations over a smaller wavelength range.}
Finally, some features ($1.079\,\mu$m and around $1.112\,\mu$m) vary in both data sets.

\begin{figure}[th]
\centering
\includegraphics[width=.4\textwidth]{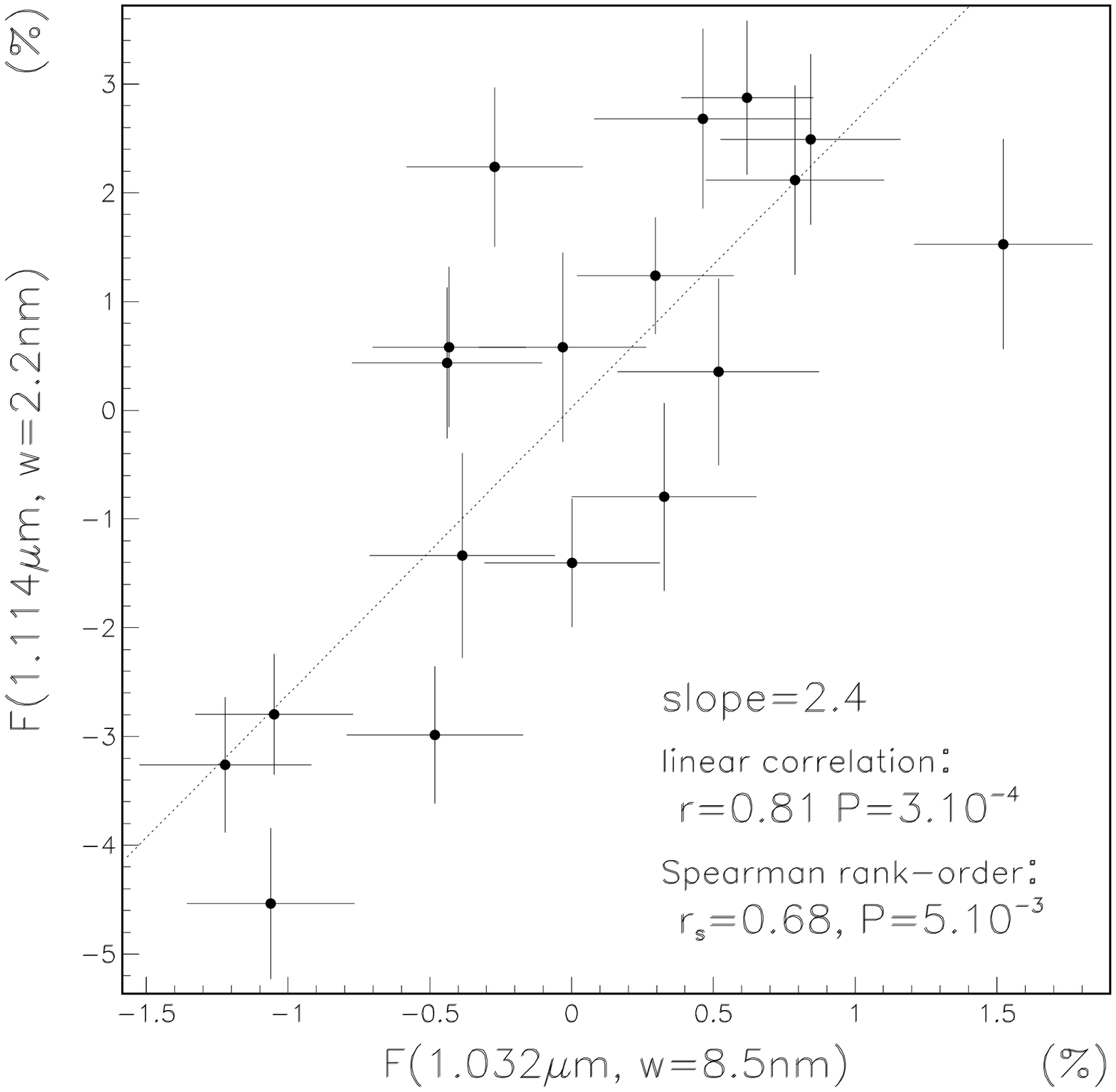}
\includegraphics[width=.4\textwidth]{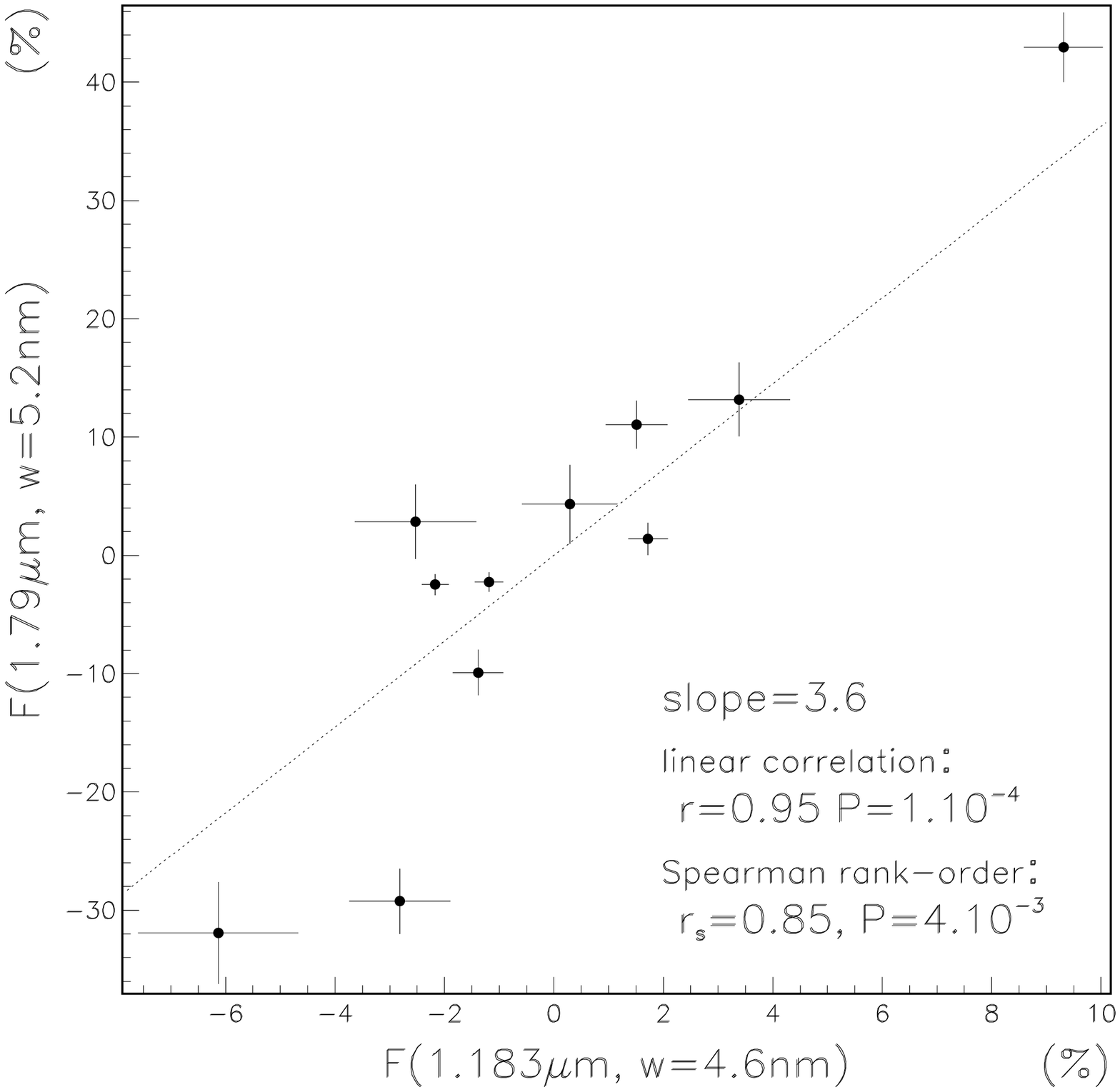}
   \caption{
     Correlation in variations for pairs of features in the spectrum of {SDSS\,J1254$-$0122},
     in the VLT spectra (top) and IRTF (bottom).
     The central wavelength and the width of the narrow bands are indicated on the axis.
     Some statistical parameters are given in the figures (particularly $P$, the probability that the variations are uncorrelated).
     For each telescope, those pairs are among the most correlated ones. 
     Ignoring any single point does not remove the correlation, except the bottom left one in the IRTF figure.
     }
      \label{fig-sd1254c}
\end{figure}

{SDSS\,J1254$-$0122} has been reported to vary in April 2002 by Goldman et~al.~(\cite{Gol03}), 
in MKO~$J, H$ and $K$ photometry, at the 5\,\% level, with no detectable colour variation, over 4\,hours.
Our IRTF spectra cannot bring constraints on these variations.

\subsection{{2MASS\,J1534$-$2952AB} (T5)}

\begin{figure}

\includegraphics[width=.5\textwidth]{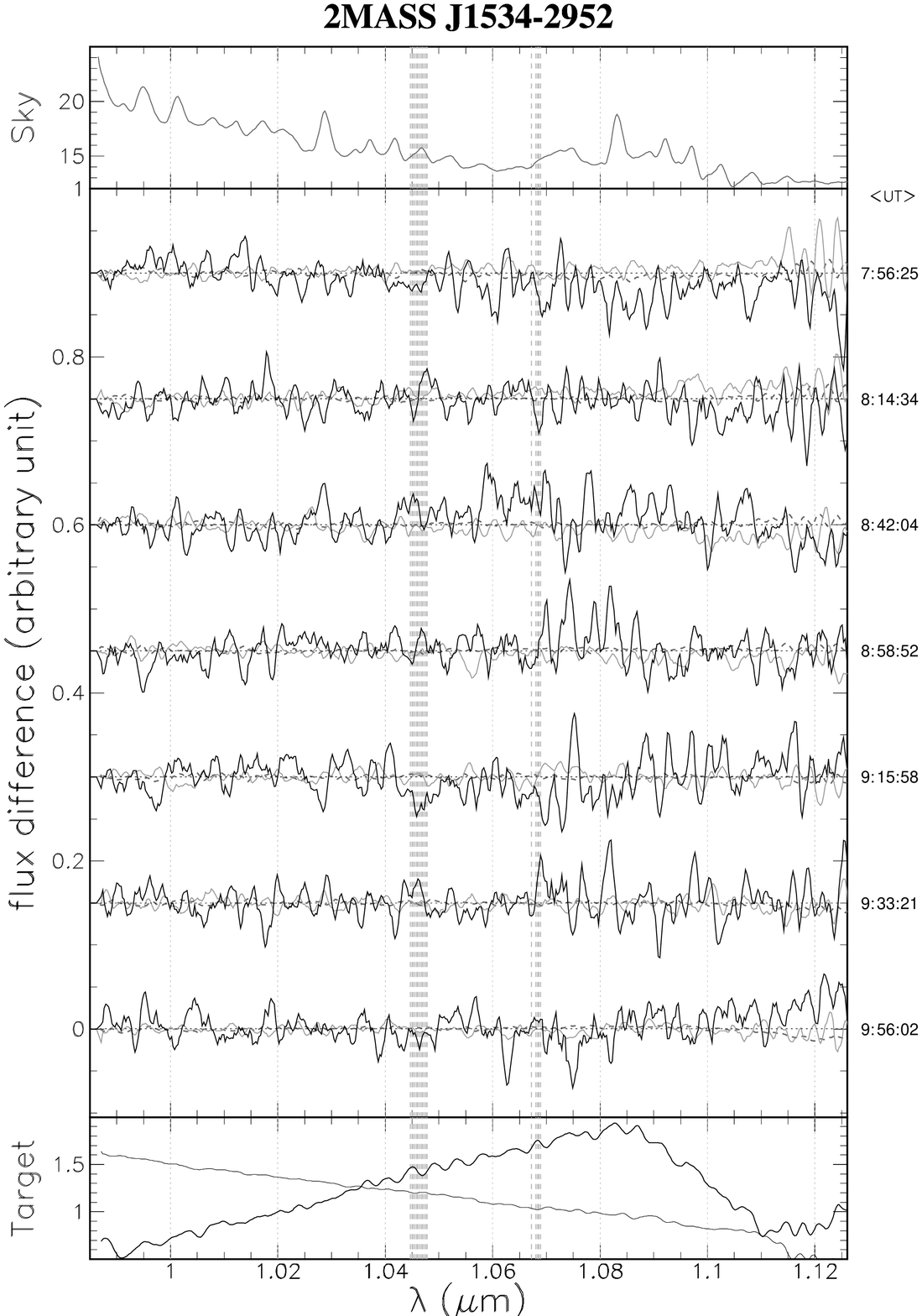}
   \caption{Spectroscopic time series for {2MASS\,J1534$-$2952AB}.
            The 15-min spectra are relative to the average of each night. 
            The flux variations of the brown dwarf (solid, black line), of the second comparison star (dashed, grey line), both relative to the bright comparison star, and the variations of the bright comparison are shown (solid, grey line).
            All spectra are normalised over the 0.997--1.054\,$\mu$m.\hspace{3cm}
            See legend of Fig.\,\ref{fig-2m0825vlt} for more details. 
         }
      \label{fig-2m1534}
\end{figure}

We obtained 46 2.5-min {\sc Isaac} spectra of {2MASS\,J1534$-$2952AB}, at the end of the night of April 16 (see Fig.\,\ref{fig-2m1534}). 
The site seeing was highly variable and increased with time {during our observations.}
The precision ranges from 1.2--2\,\% over 1.01 to~1.10\,$\mu$m to 4\,\% on the order edges and on the strong sky lines. 

By chance, a second comparison star, closer to the main comparison star, is visible in the slit, although it is not well centered. The relative variations of this second comparison star are smaller than for our target and displayed in Fig.\,\ref{fig-2m1534}. 
We only used the brighter, well centered, star to correct the fluxes of both the target and the second comparison star.

At the resolution of $R=500$, we find no features with significant variations.
At $R=100$, we find two broad features at $1.046\,\mu$m and $1.068\,\mu$m, which show variations but no correlation. 

\subsection{{2MASS\,J1225$-$2739AB} (T6)} 

\begin{figure}
\includegraphics[width=.5\textwidth]{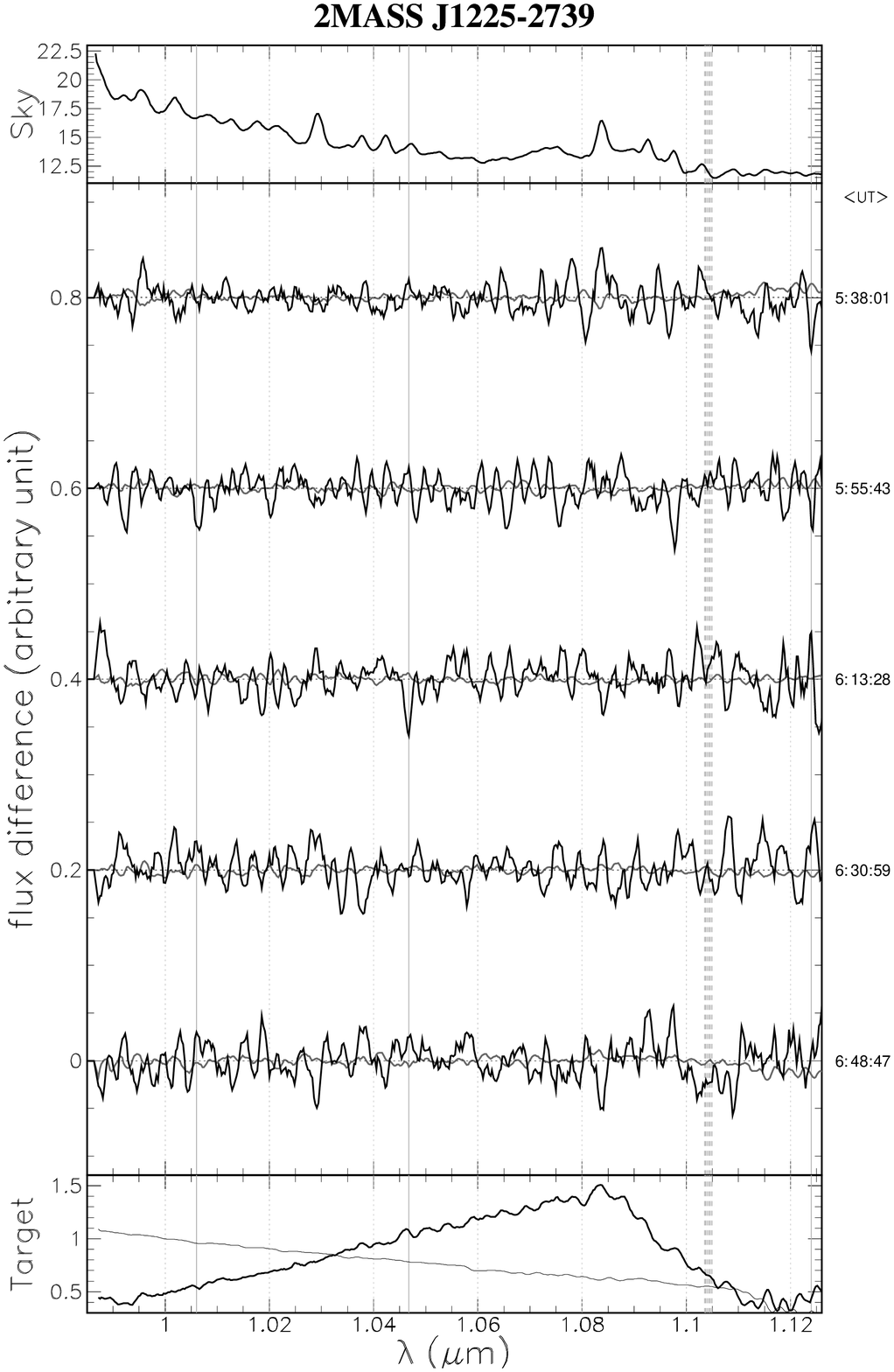}
   \caption{VLT spectroscopic time series for  {2MASS\,J1225$-$2739AB}.
            The stacked spectra are relative to the average of each night, 
            and offset for clarity. UT is the mean time since April 16, 2003, UT=0:00.
            All spectra have 15-min exposure time.
            The spectra are normalised over 1.002--1.076\,$\mu$m.
            See legend of Fig.\,\ref{fig-2m0825vlt} for details. 
           }
      \label{fig-2m1225}
\end{figure}

On April 16, we obtained a series of thirty 2.5-min {\sc Isaac} spectra (see Fig.\,\ref{fig-2m1225}), which we combined by groups of six.
The precision ranges from 1.5--2\,\% over 1.03 to~1.10\,$\mu$m to 5\,\% at 1.12\,$\mu$m. 
At the resolution of $R=500$, we find three~features with significant variations, with width of one~pixel only. 
At $R=100$, we find one broader and more interesting line at $1.104\,\mu$m, over 1.5\,nm, but due to one stacked spectrum only (see Table\,\ref{varlines}).
None of those features vary with any significant correlation with any other line.

\subsection{{SDSS\,J1624+0029} (T6)}

\begin{figure}[th!]
\includegraphics[width=.5\textwidth]{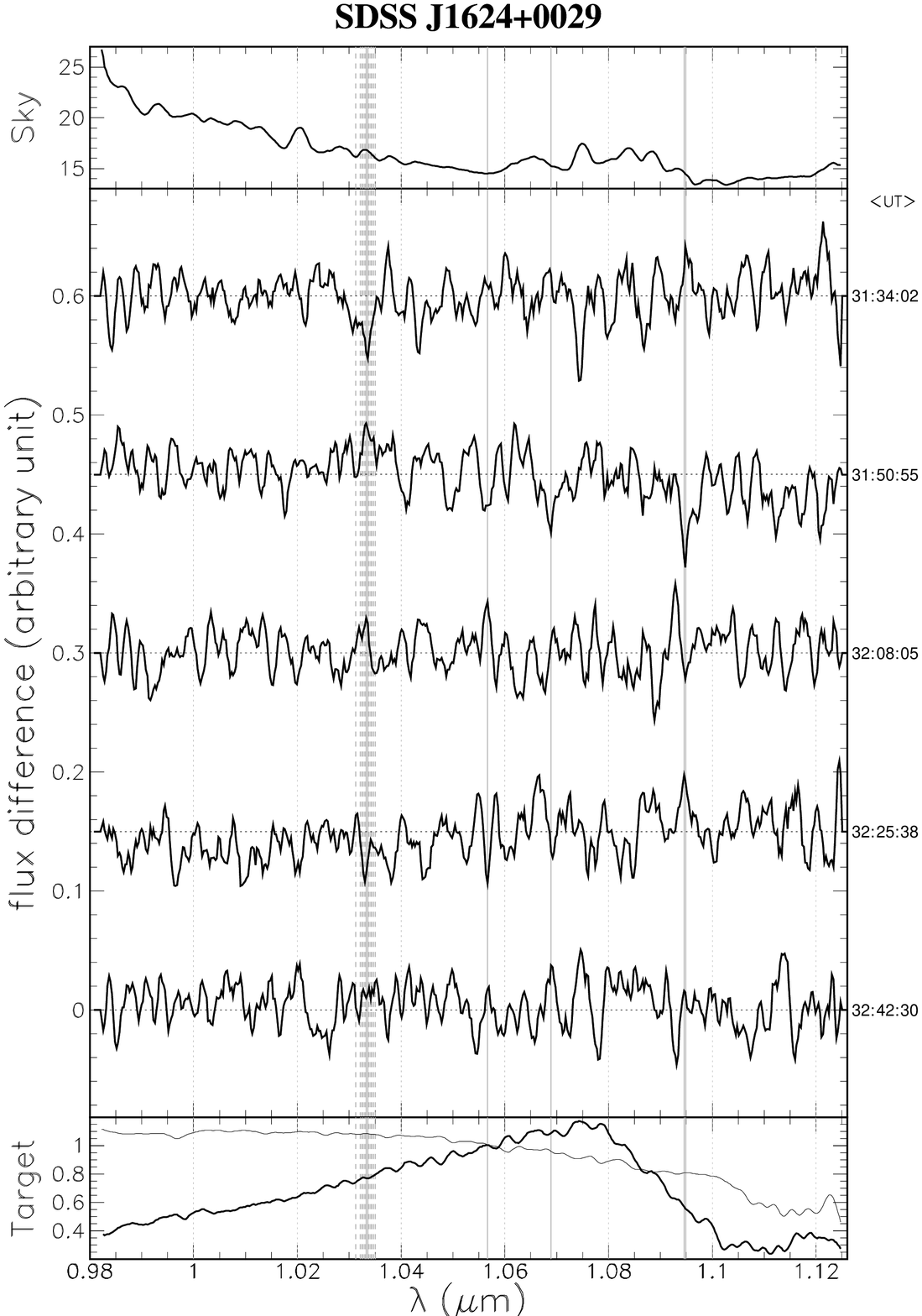}
   \caption{Spectroscopic time series for {SDSS\,J1624+0029}.
            The 15-min spectra are relative to the average of the night, and offset for clarity.
            All spectra are normalised over the 1.044--1.072\,$\mu$m region.
            See legend of Fig.,\ref{fig-2m0825vlt} for more details. 
      }
      \label{fig-sd1624}
\end{figure}

We obtained 42 2.5-min {\sc Isaac} spectra of {SDSS\,J1624+0029}, at the end of the night of April 17,
which we stacked by groups of six (see Fig.\,\ref{fig-sd1624}).
The precision ranges from 2\,\% over 1.03 to~1.09\,$\mu$m to 5\,\% on the order edges and on the strong sky lines. 
The last 12~spectra are discarded due to high extinction and {slit losses} (totalling 1\,mag).
At the resolution of $R=500$, we find four features with significant variations, of width one or two pixels, which show no correlation.
At $R=100$, we find one broad line at $1.033\,\mu$m with width 5\,nm.
There is a FeH absorption line at $1.03398\,\mu$m  (Cushing et~al.~\cite{Cus03});
for such a cool object however FeH is not expected to play a major role, and we do not see variations in the stronger 0.99-$\mu$m band head.

Nakajima et~al.~(\cite{Nak00}) reported a variation in the water band at 1.53--1.58\,$\mu$m using the Subaru telescope,
although over only 80\,min of time. 
It is difficult to estimate the amplitude and significance of the variability in their data.
We do not see significant variations at the edge of the water band absorption at $1.1\,\mu$m, in our {\sc Isaac} spectra.

\section{Discussion} \label{discussion}

\subsection{Models of cloud fragmentation} \label{model}

\begin{figure*}
\includegraphics[width=\textwidth]{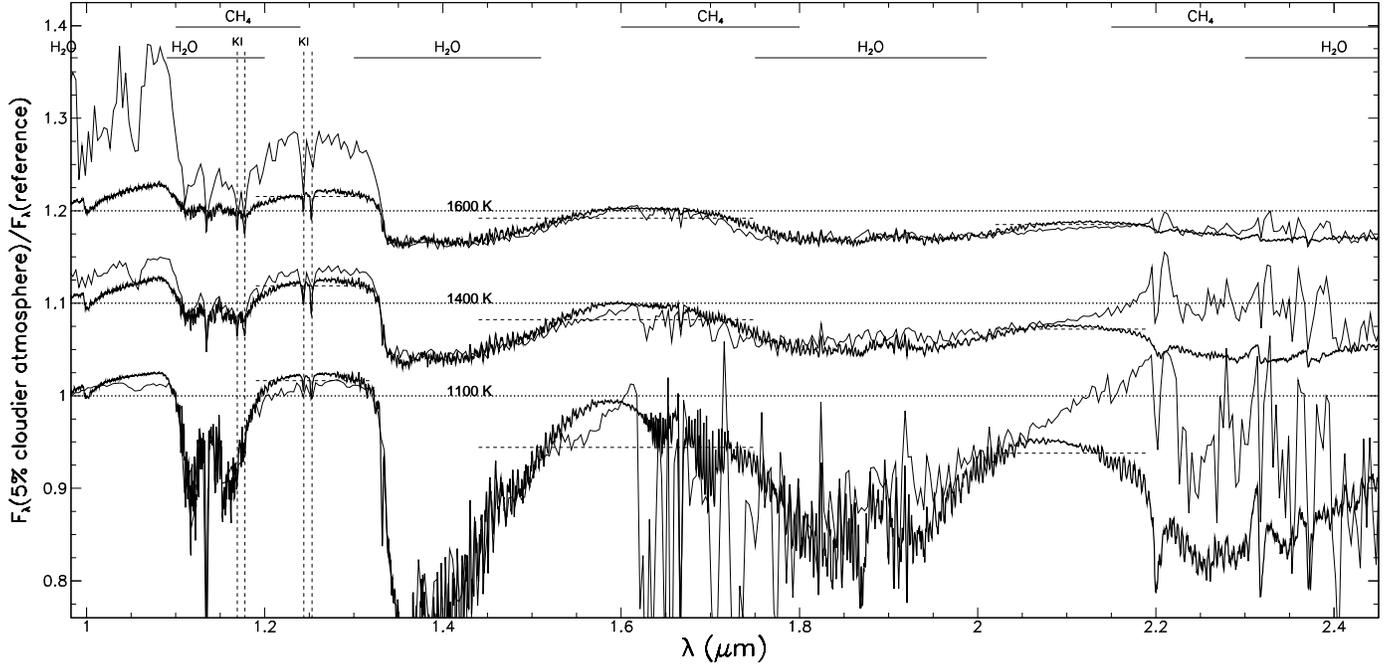}
   \caption{Crude model predictions for the IRTF wavelength range.
            We used the $f_{sed}=3$ and {\tt clear} models (Marley et~al.~\cite{Mar02}, in thick lines) and {\tt Cond}, and {\tt Settl05} and {\tt Cond02} models (Allard et~al, priv.com., in thin lines).
            The model spectrum resolutions are degraded to roughly match the {\sc SpeX} actual resolution.
            See legend of Fig.\ref{fig-model-VLT} for details.
            Differences between the models in the water bands 
            are likely due to the different assumptions regarding the cloud opacity structure.
            Horizontal dashes lines indicate the wavelength range and effect of the normalisation in the $J$, $H$ and $K$ bands.
           }
      \label{fig-model-IRTF}
\end{figure*}

\begin{figure}
\includegraphics[width=.5\textwidth]{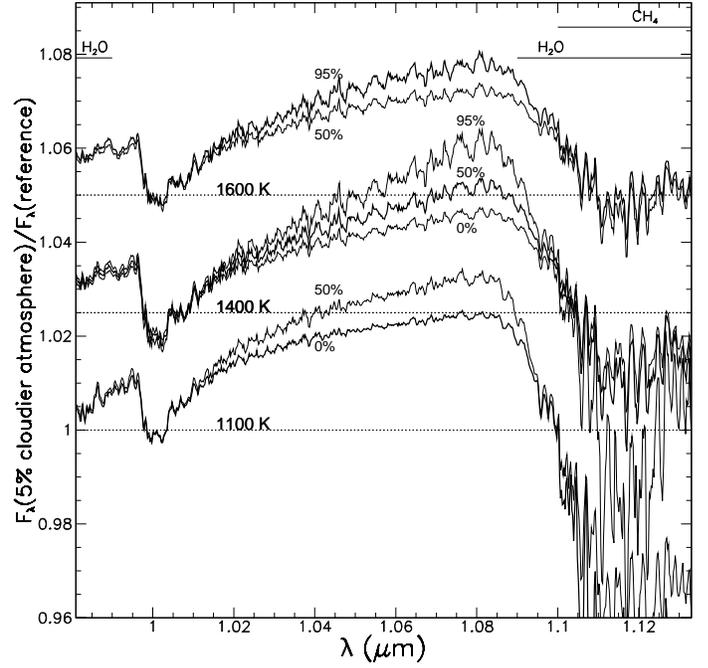}
   \caption{Crude model predictions for the VLT wavelength range, for three effective temperatures: 1600, 1400 and 1100\,K from top to bottom, for a gravity of $10^3\,{\rm m/s}^2$.
           We used the $f_{sed}=3$ and {\tt clear} models (Marley et~al.~\cite{Mar02}).
            The model spectrum resolutions are degraded to roughly match the {\sc Isaac} actual resolution.
            We show the ratio of the model spectra with clouds covering 5\,\% more of their disk compared to those expected for each effective temperature {(thick lines) and other cloudy- to total area percentages (as indicated in the graph)}.
             The amplitude of the computed variation depends linearly on the assumed fractional change of cloud coverage. 
      }
      \label{fig-model-VLT}
\end{figure}

{One of the proposed explanations} to the reported brown dwarf variability 
and the rapid disappearance of the cloud deck at the L/T transition
is the fragmentation of the cloud deck (Ackerman \& Marley~\cite{Ack01}; Burgasser et~al.~\cite{Bur02b}).
{This is only one of several hypothetical mechanisms to produce variability. }
We compare our results 
with crude theoretical predictions
based on 
that hypothesis.
We assume that the flux coming from a partly cloudy brown dwarf is 
the sum of the flux predicted by either the {\tt Cloudy} and {\tt Clear} or {\tt Settl} and {\tt Cond} models,
weighted by the cloud and clear coverage.  Here {\tt Cloudy} and {\tt Settl} are the
vertically constrained cloudy atmosphere models by Ackerman \& Marley~(\cite{Ack01}) and Allard et~al.~(\cite{All03}, 2006, priv.com.), respectively. 
{\tt Clear} and {\tt Cond} are the respective cloud-free models, which have
similar, but not identical, assumptions built into their construction.  
This simulation should only be taken as a guide, in the absence of dynamic models.
In Fig.\,\ref{fig-model-VLT} and \ref{fig-model-IRTF}, we show the {\it relative} effect of the appearance of a 5\,\% cloud 
(replacing a cloud-clear region) on a brown dwarf.
{The set of temperatures and parameters we use are intended to be representative of our targets rather than the best fitted values for each of our five targets.
We estimate that differences between our predictions due to change in substellar parameters are smaller than the uncertainties of the predictions themselves.}
For each temperature, we adopt as a reference cloud coverage expected by the atmosphere models.
Hence,  a 1\,100\,K brown dwarf is well reproduced by clear model, so we consider the spectroscopic variations introduced by the appearance of a 5\,\% cloud on a clear brown dwarf.
Similarly a 1\,600\,K brown dwarf is well described by a cloudy model, and we show the spectroscopic variations introduced by the disappearance of a 5\,\% cloud on a fully cloudy brown dwarf.
For the 1\,400\,K brown dwarf, such as {SDSS\,J1254$-$0122}, we compare a 55\,\% covered atmosphere to a 50\,\% one.
{How does the actual ratio of cloudy to clear areas change the expectations? 
The importance of using different reference cloud coverages are illustrated in Fig.\ref{fig-model-VLT}.
A similar study over 1--2.5$\,\mu$m shows the same trend, except in the water bands (1.4$\,\mu$m, 1.9$\,\mu$m) where we have little or no signal.}
The {\it relatively} larger variations seen in the water band of cooler brown dwarfs are due to the smaller flux escaping at those wavelengths.

In the VLT 0.99--1.13\,$\mu$m spectral region, this implementation of the Ackerman \& Marley~(\cite{Ack01}) models predicts {narrow-band} variations of small (unndetectable) amplitude,
while larger slope variations are predicted over 1.0--1.08\,$\mu$m and over 1.08--1.115\,$\mu$m.
Large  {narrow-band} variations, arising from the strong water bands, are expected redward of 1.12\,$\mu$m,
but are difficult to observe from the ground.
In the $J$, $H$ and $K$ bands, Marley et~al.~(\cite{Mar02}) predict strong variations in the water bands,
where our spectra are too noisy to draw conclusions.
Variations similar to those of the 1.0--1.1\,$\mu$m band are expected in the $J$~band, 
and {anti-correlated} variations are expected in the $K$~band.  Comparison of Fig.\,\ref{fig-model-IRTF} to Fig.\,7 of
Ackerman \& Marley~(\cite{Ack01}) gives a qualitative interpretation of the origin these effects.

From these calculations we can deduce an estimate for the expected variations: 
$\sim$3\,\% broad-band flux increase over 0.99--1.13\,$\mu$m and in the $J$~band when a 5\,\% clear region is replaced by a cloudy region.
{According to our application of the Marley et~al.~models, narrow-band variations in the {\sc Isaac} wavelength range and in $J$~band, except for {\sc K~i}, are not predicted.
In the $H$ and $K$~bands, and in Allard et~al.~models, the situation is more complex. 
}
For  {2MASS\,J1534$-$2952AB}, the observing conditions {(high winds)} and flux losses due to the slit require to normalise of the spectra before searching for variations (see Section\,\ref{red}).
A similar normalisation is performed on {all observed spectra (for homogeneity) and} the synthetic spectra, so that we are only sensitive to variations of the slope.

\begin{figure}
  \includegraphics[width=.5\textwidth]{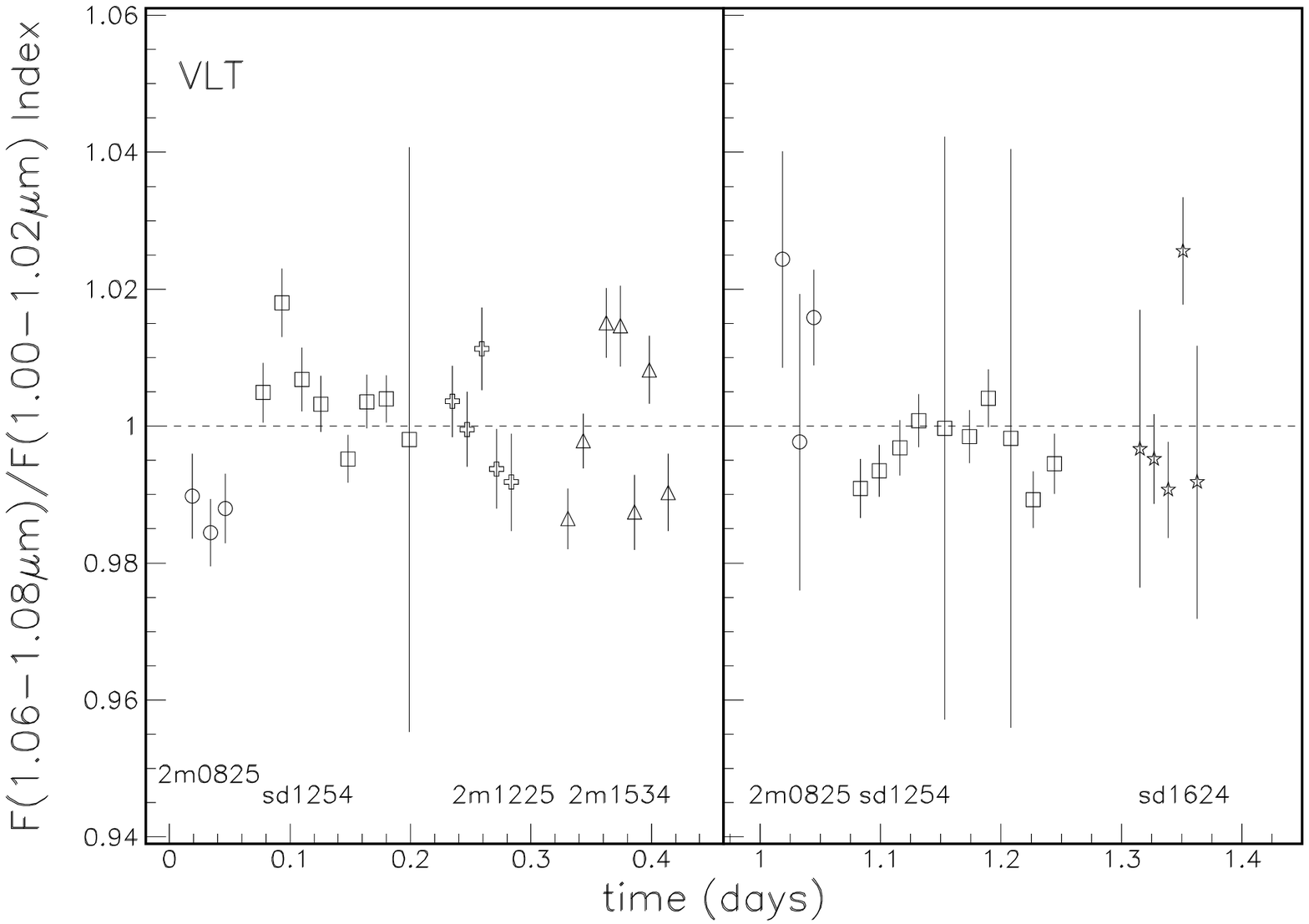}
  \includegraphics[width=.5\textwidth]{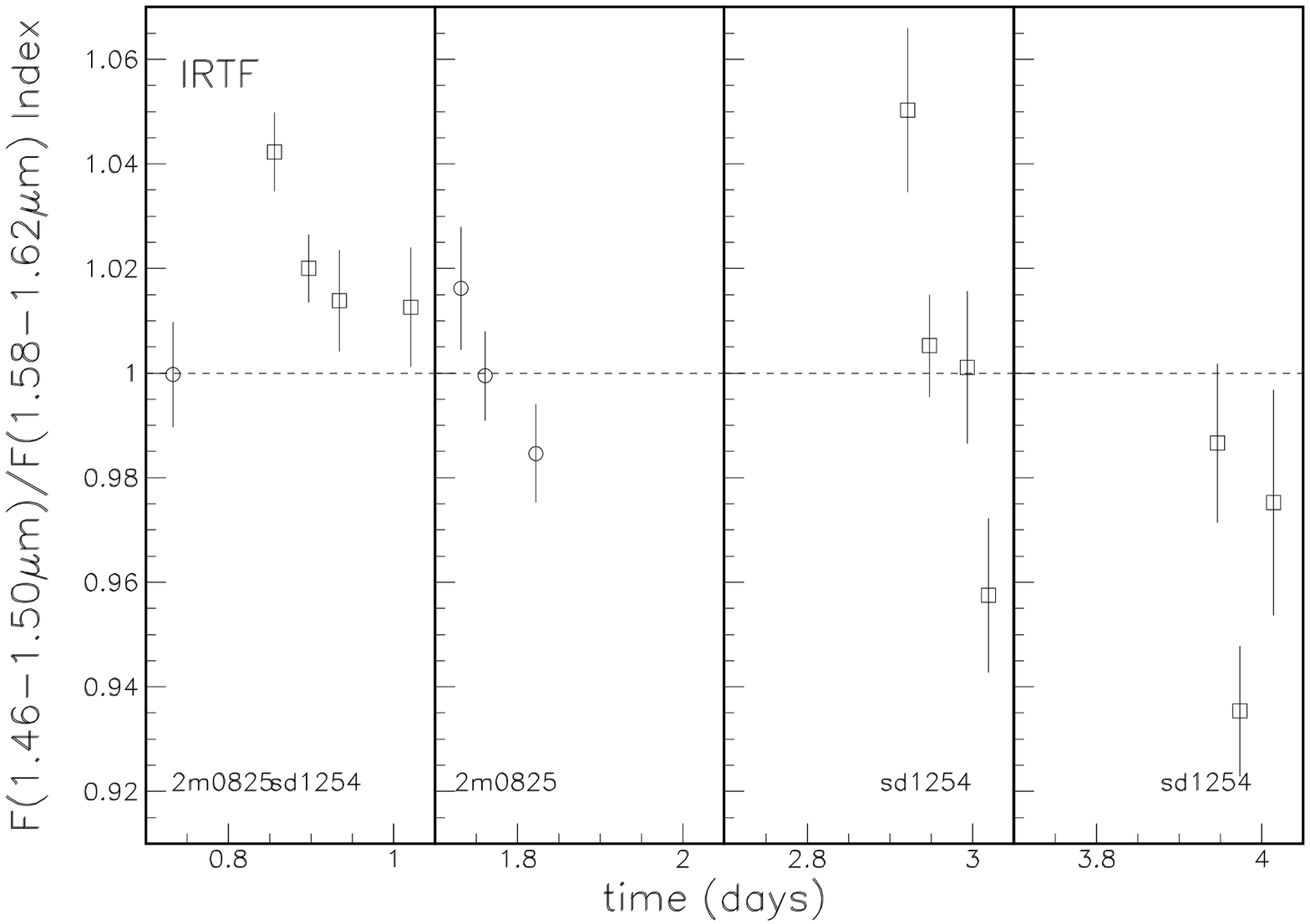}
  \caption{Spectral index from the VLT (top) and IRTF (bottom) data.
  For clarity, the indices are normalised to 1 on average, independently for each target. (The normalisation factor is primarily a function of the spectral type.)
  Target names are indicated.
  }
  \label{fig-index}
\end{figure}

{We therefore calculate the variations of a spectral index, 
$$s_{\rm VLT}=\frac{1.00-1.02\,\mu\rm m}{1.06-1.08\,\mu\rm m},$$
for the data and synthetic spectra, after normalisation. 
The spectral index is chosen in order to maximize the predicted variability signal, as illustrated by the Fig.\,\ref{fig-model-VLT}, and minimize the photo-noise.
However, small changes in the definition of the index (shift by $\pm 5$\,nm) have negligible effects on the final results, up to 15\,\% on the observed dispersion of the index (see Fig.\ref{fig-index}) and on the variability upper-limits (see Table\,\ref{tab-index}) we derive.
As discussed earlier, reliable errors are necessary to properly calculate the upper-limits. 
Here we assumed that the noise from neighbouring pixels are not correlated, as is expected for photo-noise, and we simply square-add errors of individual pixel fluxes.

This also applies to the IRTF observations.
As a result, the variations we are really sensitive to, for the same cloud-cover variations, are now small in the $J$~band, except at the {\sc K~i} doublet feature.
In the $H$~band, the main variations are due to water and methane absorptions at the edges of the band, mostly affecting the broad shape of the spectrum over that wavelength range.
In the $K$~band, variations are detected only for the coolest temperature (1\,100\,K) at the red end of the band, and are due to variable methane absorption.
Those expectations are however strongly model-dependent.

For the IRTF data we use the spectral index: 
$$s_{\rm IRTF}=\frac{1.46-1.50\,\mu\rm m}{1.58-1.62\,\mu\rm m},$$ 
which again maximizes our sensitivity while avoiding the main telluric water bands.
}

\begin{table}
  \caption[]{Dispersions of VLT and IRTF spectral indices, upper-limits (at 99\,\% C.L.) calculated from the observed errors, and expected variations due to cloud cover change assuming our simple model and a  cloudy area increase of 5\,\% of the dwarf surface. The models are characterized by their effective temperature and the cloud coverage, and are based on the Ackerman \& Marley~(\cite{Ack01}) grid of models.
  }
  \label{tab-index}
  \begin{tabular}{lcccr@{, }r}
    \hline
    \hline
    \noalign{\smallskip}
    Target               & $\sigma(s)$              &  Upper-lim. & Expected & \multicolumn{2}{c}{Model}  \\
    \noalign{\smallskip}
    \hline
    \noalign{\smallskip}
   VLT        \\
    2MASS\,J0825 &  1.5\,\%  & 2.2\,\%   & 2.1\,\%  &  1\,600\,K& 95\,\%        \\ 
    SDSS\,J1254    &  0.7\,\%  & 1.1\,\%   & 2.2\,\%  &  1\,400\,K& 50\,\%        \\ 
                                &               &              & 2.9\,\%  &  1\,400\,K& 95\,\%        \\
                                &               &              & 1.9\,\%  &  1\,400\,K& 5\,\%        \\
    2MASS\,J1225 &  0.7\,\%  & 2.2\,\%   & 1.7\,\%  &  1\,100\,K& 0\,\%        \\ 
    2MASS\,J1534 &  1.2\,\%  & 1.6\,\%   & 1.7\,\%  &  1\,100\,K& 0\,\%        \\ 
    SDSS\,J1624    &  1.3\,\%  & 2.7\,\%   & 1.7\,\%  &  1\,100\,K& 0\,\%        \\ 
    \noalign{\smallskip}
    \hline
    \noalign{\smallskip}
   IRTF        \\
    2MASS\,J0825 & 1.1\,\%   &  2.1\,\%  & 2.6\,\%  &  1\,600\,K& 95\,\%        \\ 
    SDSS\,J1254   & 3.3\,\%   &  1.8\,\%  & 4.1\,\%  &  1\,400\,K& 50\,\%        \\ 
    \noalign{\smallskip}
    \hline
  \end{tabular}
\end{table}

\subsection{Interpretation} \label{interpretation}

The rotational velocity of {2MASS\,J0825+2115} was measured by Bailer-Jones~(\cite{CBJ04}) {\rm to be $16.9^{+4.5}_{-5.6}\,$km/s} and corresponds to a rotational period between 2.48 and 10.75\,h (with 90\,\% of confidence levels, 
which include the rotation axis orientation uncertainty and conservative measurement errors). 
Hence the observing time of each night only samples part of one period, while the 2003, April~16 and 17 VLT data may have been taken one to nine~periods apart.
{The IRTF data, collected over 50\,hours, also cover several rotational periods.}
{Zapatero Osorio et~al.~(\cite{ZO06}) similarly measured $v \sin i=27.3\pm2.5$\,km/s and $38.5\pm2.0$\,km/s for {SDSS\,J1254$-$0122} and {SDSS\,J1624+0029} respectively, corresponding to a period range about twice as short as that of {2MASS\,J0825+2115}.
Therefore, each night of observations most probably covers half a rotational period or more of {SDSS\,J1254$-$0122} and {SDSS\,J1624+0029}.
{2MASS\,J1225$-$2739AB} and {2MASS\,J1534$-$2952AB} have no} measured rotational velocity. 
However, brown dwarf period estimates are all in the range of 1.5 to~13\,hours (Basri et~al.~\cite{Bas00}; Mohanty \& Basri~\cite{Moh03}; Bailer-Jones~\cite{CBJ04}, Zapatero Osorio et~al.~\cite{ZO06}), 
so rotational periods for our targets are likely to fall in this range of values.
If the brown dwarf's surface is a patchwork of clouds and cloud holes, 
the cloud cover ratio might change with time as the brown dwarf rotates.
Alternatively, dynamical processes could alter the ratio of cloudy to clear regions.
{In the latter case, little is known about the variability time scale to expect.}

\begin{figure}
\centering
\includegraphics[width=.5\textwidth]{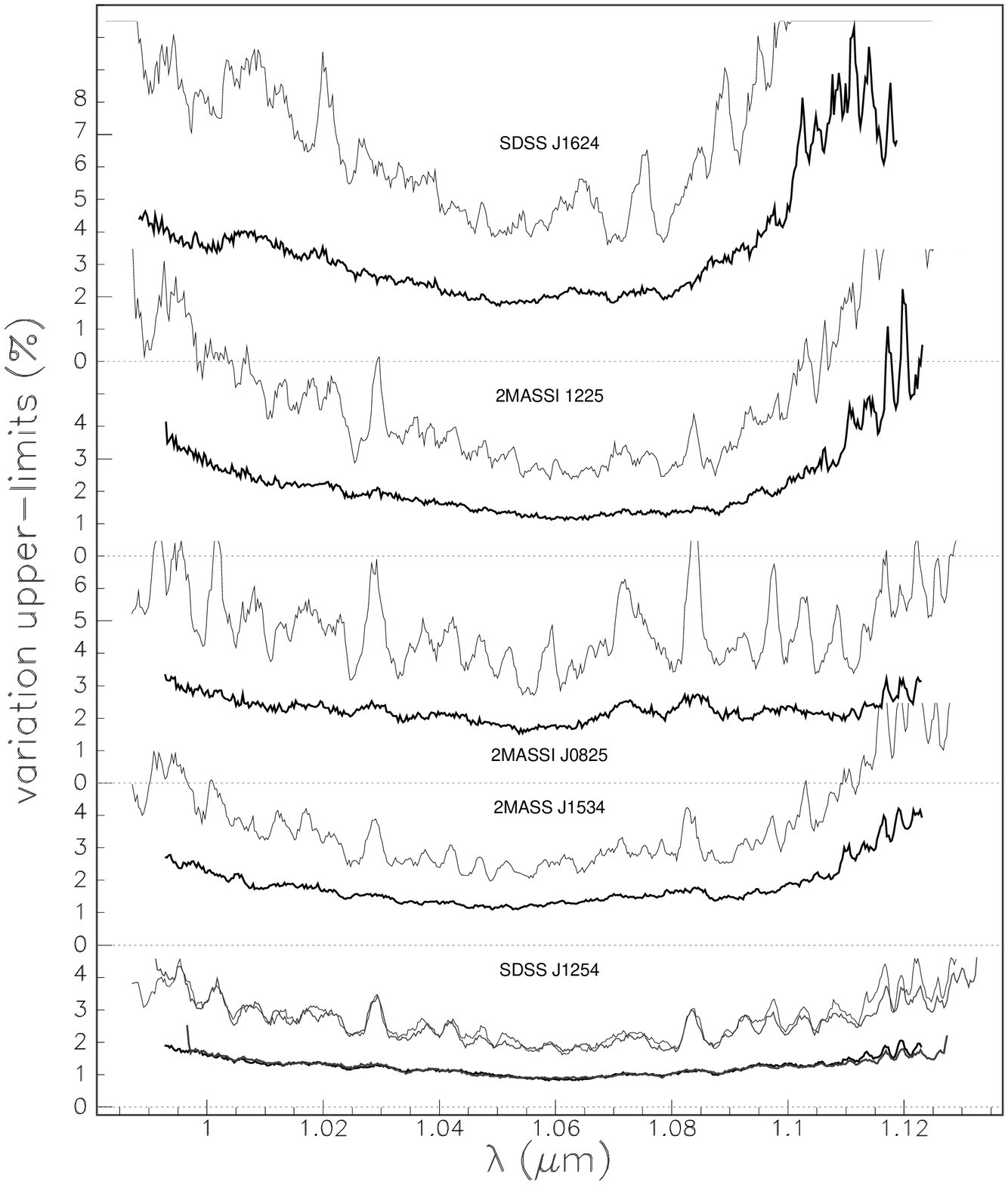} 
   \caption{ 
     Variation upper limits for the VLT sample, relative to the night average ({99\,\%-confidence-level}),
     after smoothing over 11\,pixels or 3\,nm (thin lines), or over 51\,pixels or 15\,nm (thick lines).
     For {SDSS\,J1254$-$0122} (almost identical) independent results from both nights are shown;
     for {2MASS\,J0825+2115}, significant differences appear, due to the different weather conditions, and we represent the upper-limits mean of the two nights.
     Large upper-limits (redward of $1.1\,\mu$m for {2MASS\,J1225$-$2739AB} and {SDSS\,J1624+0029}) are truncated for clarity.
   }
      \label{fig-upperlim-vlt}
\end{figure}

\begin{figure*}
\centering
\includegraphics[width=\textwidth]{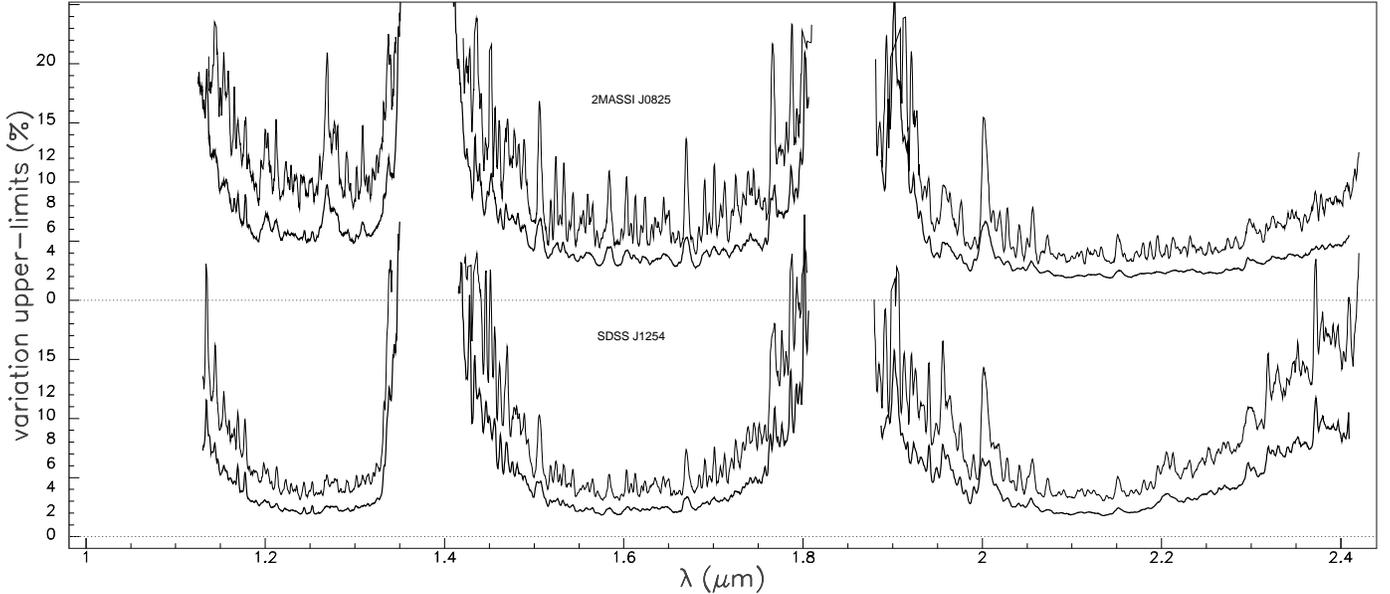} 
   \caption{ 
     Variation upper limits for the IRTF sample, relative to the average ({99\,\%-confidence-level}). 
     Large upper-limits (in the water bands) are truncated for clarity.
     In contrast to Fig.\,\ref{fig-upperlim-vlt}, upper-limits are calculated for the whole run, not on a nightly basis.
   }
      \label{fig-upperlim-irtf}
\end{figure*}

{We do not see the broad-band variations that 
would be expected if there are large fractional changes in global cloud cover,
using both the Ackerman \& Marley~(\cite{Ack01}) and Allard et~al.~(\cite{All03}, 2006) models for the 1800--1100\,K effective temperature range.
}
We show our variability upper limits, for broad {and narrow bands}, in Fig.\,\ref{fig-upperlim-vlt} and \ref{fig-upperlim-irtf}.

{We can use our spectral indices to draw quantitative conclusions. 
In Table\,\ref{tab-index}, we report the observed dispersions and 99\,\%-confidence-level upper-limits, as well as the variations expected based our application of the the Ackerman \& Marley~(\cite{Ack01}) models,
}
for a change of 5\,\% in the cloud cover, for the effective temperatures of interest.
{Thus, with upper-limits on the broad-band variability over 0.99--1.13\,$\mu$m 
of 1.1--2.7\,\%, depending on the targets,
we can exclude {\it global} changes larger than $\sim$5--8\,\% in the ratio of cloudy to clear regions,
{during the time we observed the brown dwarfs} for four of our targets.
The constraint based on the {\sc SpeX} data for {2MASS\,J0825+2115} in the $H$~band is 4\,\%.}
We cannot exclude the possibility that many changing regions in the cloud cover, each of smaller typical scale, affect a large surface of the brown dwarfs.

{Some of the spectroscopic time series do exhibit high-spectral frequency variability, 
which we report in Table\,\ref{varlines}.
Generally, the amplitudes of the variations are small compared to the noise, so that we cannot confidently claim that these variations are {real and intrinsic to the source}. 
The low-amplitude variations observed at 0.997\,$\mu$m in the spectra of {SDSS\,J1254$-$0122} and {2MASS\,J0825+2115}, and at 1.033\,$\mu$m for {SDSS\,J1624+0029}, fall within an FeH band. However we detect no variations at the 0.99\,$\mu$m band head as may be expected if FeH absorption variations are responsible for the 0.997\,$\mu$m variability.
A search for variability at other wavelengths at which the object is bright and FeH is a major absorber may be indicated, such as the complex of FeH features in $H$~band (Cushing et~al.~\cite{Cus03}). 
{In the optical, such variations might have been detected around the CrH and FeH bandheads between 8600 and 8700\,\AA\ (Kirkpatrick et~al.~\cite{Kir01}).}
Since different bands probe different depths in the atmosphere, 
variations between the amplitude of variability in the FeH bands 
could reveal the underlying source of the variability we observe, if it is indeed real.
It should be noted that some low-amplitude variations are detected in the spectra of several targets:
Around 1.03\,$\mu$m for {SDSS\,J1624+0029} and {SDSS\,J1254$-$0122},
or 1.046\,$\mu$m for {2MASS\,J1534$-$2952AB} and {SDSS\,J1254$-$0122}.
As we used different flat-field and the calibration slightly varies, we do not expect an instrumental effect.
However we cannot relate those features with a known absorber.
}
For all targets except {SDSS\,J1254$-$0122}, we do not consider that these variations impair our conclusions regarding the maximal cloud size.
  
\begin{table*}
  \caption[]{
Partial list of features which we report to vary with a confidence level better than 99\,\%. 
We only report the most significant, or most robust varying features.
We indicate in the third column additional features whose variations are correlated in time (at 99.9\,\% C.L.).
  } 
  \label{varlines}
  \begin{tabular}{lcccrc}
    \hline
    \hline
    \noalign{\smallskip}
    Target   
                 & telescope & wavelength & width & peak-to-peak & correlated features \\
                 &                     & ($\mu$m)     &  (nm) & &  ($\mu$m)   \\
    \noalign{\smallskip}
    \hline
    \noalign{\smallskip}
    {2MASS\,J0825+2115} (L6)      
      & VLT & 0.996 & 1   & 11\,\% & \ldots \\
      & VLT & 1.008 & 5   &   5\,\% & \ldots \\
      & VLT & 1.065 & 5   & 14\,\% & \ldots \\ 
    {2MASS\,J1225$-$2739AB} (T6) 
      & VLT & 1.104 & 1.5 & 6\,\% & \ldots \\
    {SDSS\,J1254$-$0122} (T2) 
      & VLT & 0.997 & 1.2 & 6\,\% & \ldots \\
      & VLT & 1.031 & 3  & 4\,\% & 1.053, 1.103, 1.115 \\ 
      & VLT & 1.046 & 4 & 6\,\% & \ldots \\
      & VLT & 1.12 & 20 & 10\,\% & \ldots \\
      & IRTF & 1.127 & 4 & 60\,\% & 1.472, 1.764 \\ 
      & IRTF & 1.471 & 3 & 45\,\% & 2.324 \\
      & IRTF & 1.580 & 5 & 14\,\% & 2.062 \\
      & IRTF & 1.617 & 3 & 25\,\% & 2.247, 2.281 \\ 
      & IRTF & 2.062 & 4 & 22\,\% & 2.415 \\
      & IRTF & 2.246 & 5 & 45\,\% & 2.410 \\
      & IRTF & 2.281 & 5 & 50\,\% & 2.324 \\ 
    {2MASS\,J1534$-$2952AB} (T5) 
      & VLT & 1.046 & 4 & 4\,\% & \ldots \\
      & VLT & 1.068 & 6 & 7\,\% & \ldots \\
    {SDSS\,J1624+0029} (T6)       
          & VLT & 1.033 & 6 & 7\,\% & \ldots \\
    \noalign{\smallskip}
    \hline
  \end{tabular}
\end{table*}

\subsection{The case of SDSS\,J1254$-$0122} \label{sd1254interp}

The case of {SDSS\,J1254$-$0122} is different, as in both data set, its spectrum exhibits numerous significant variations.
A subset is common to both instruments, and a significant proportion {of variable lines} show correlated variations.
For this object we therefore conclude that the detected variations are possibly real.
Some, but not all, of the KI doublet lines are affected, as well as other lines which our simple models suggest could be variable. 
However, other features in the continuum are also seen to vary, which the simple model does not predict.  
We note that many of the variable features are found in region of the spectrum affected by telluric water band absorption. {Although they sometimes correlate with each other, there is no systematic correlation}, and other variable features are found where there are little or no telluric absorption.
{We also point out that some highly variable features detected in the SpeX data fall close the order edges or in regions of small SNRs, where our error determination may become inaccurate.}

{Among our targets, {SDSS\,J1254$-$0122} represents the prototypical case of a transition object.
It is therefore worth noting that it shows higher levels of variability than our other four targets, and is possibly the only one that we could classify as variable.
This is somewhat counterbalanced by the facts that we obtained more good-quality data on that object, and that it is the brightest (along with {2MASS\,J1534$-$2952AB}). 
So that our sensitivity is higher.
Finally, the small size of our sample precludes any conclusion regarding the variability frequency among L/T transition objects, in comparison with earlier and later brown dwarfs.  Nevertheless we suggest additional monitoring of this object.
 }

\section{Conclusions}

  We have obtained high SNR, low resolution spectroscopic time series over 0.99--1.13\,$\mu$m
  for five brown dwarfs of spectral types L6 to T6, sampling the L/T transition,
  and for a sub-set of two brown dwarfs of types L6 and T2 in the $J, H$ and $K$ bands.
  We cautiously report some high-frequency {(narrow-band)} variations, 
  which we generally cannot tie to specific absorption lines.
  {An exception is FeH in the spectra of three targets.}
  {SDSS\,J1254$-$0122} shows numerous variable features.
  All require confirmation by additional observations.

  We place constraints on broad-band spectroscopic variability 
  at the levels of 2\,\% ({2MASS\,J1534$-$2952AB}) to
  3\,\% ({2MASS\,J0825+2115}, {2MASS\,J1225$-$2739AB} and SDSS\,J1624+0029),
  {with a 99\,\% confidence level.}

  When comparing to the variations expected from crude atmospheric model interpolations
  between cloudy, L-type and clear, T-type atmospheres,
  we find that over the course of our observations, 
  no significant variations (larger than $\sim$5--8\,\% of the brown dwarf's disk)
  in the cloud coverage occurred in our sample.
  {Our data do not rule out smaller-size heterogeneity on the brown dwarf surface, nor does our analysis try to constrain other variability mechanisms.} 

\begin{acknowledgements}
       B.~Goldman thanks F.\,Clarke and the ESO staff for its support during and after his VLT run, as well as France Allard for valuable discussions.
       B.~Goldman and M.~Marley acknowledge support from NASA grants NAG5-8919 and NAG5-9273.
       M.~Cushing acknowledges financial support from the NASA Infrared Telescope Facility.
       A.~Burgasser kindly provided unpublished finding charts for some of our targets.
       This Research has made use of the M, L, and T dwarf compendium housed at DwarfArchives.org and maintained by C. Gelino, D. Kirkpatrick, and A. Burgasser, 
       and of the {\sc Simbad} database, operated at C.D.S., Strasbourg, France.
       Visiting Astronomer at the Infrared Telescope Facility, which is operated by the University of Hawaii under Cooperative Agreement no. NCC 5-538 with NASA, Office of Space Science, Planetary Astronomy Program.
\end{acknowledgements}


\begin{thebibliography}{}

  \bibitem[2001]{Ack01} Ackerman, A. S. \& Marley, M. S.
                        2001, ApJ, 556, 872 
                        
  \bibitem[2003]{All03} Allard, F., Guillot, T., Ludwig, H.-G., et~al.
                        2003, Brown Dwarfs, IAU Symposium \#211. Edited by E.~Mart\'\i n. San Francisco: PASP, 325

  \bibitem[2000]{Bas00} Basri, G., Mohanty, S., Allard, F. et~al.
                        2000, ApJ, 538, 363 

  \bibitem[2004]{CBJ04} Bailer-Jones, C. A. L. 
                        2004, A\&A, 419, 703 

  \bibitem[2008]{CBJ08} Bailer-Jones, C. A. L. 
                        2008, MNRAS, {\it in press}, {\tt astro-ph/0711.4464} 

  \bibitem[2003]{CBJ03} Bailer-Jones, C. A. L. \& Lamm, M. 
                        2003, MNRAS, 339, 477
                        
  \bibitem[1999]{CBJ99} Bailer-Jones, C. A. L. \& Mundt, R.
                        1999, A\&A, 348, 800

  \bibitem[2001]{CBJ01} Bailer-Jones, C. A. L. \& Mundt, R.  
                        2001, A\&A, 367, 218 \&
                        2001, A\&A 374, 1071 (erratum)
                       
  \bibitem[2003]{Bou03} Bouy, H., Brandner, W., Mart\'\i n, E. L., et~al.
                          2003, AJ, 126, 1526
                        
 \bibitem[1999]{Bur99} Burgasser, A. J., Kirkpatrick, J. D., Brown, M. E., et~al.
                        1999, ApJ, 522, L65 

 \bibitem[2002a]{Bur02a} Burgasser, A. J., Kirkpatrick, J. D., Brown, M. E., et~al.
                        2002a, ApJ, 564, 421 
                        
 \bibitem[2002b]{Bur02b} Burgasser, A. J., Marley, M. S., Ackerman, A. S., et~al.
                        2002b, ApJ, 571, L151 

 \bibitem[2002c]{Bur02c} Burgasser, A. J., Liebert, J., Kirkpatrick, J. D., \& Gizis, J. E., 
                        2002c, AJ, 123, 2744 

  \bibitem[2003]{Bur03} Burgasser, A. J., Kirkpatrick, D. J., Reid, I. N., et~al.
                        2003, ApJ, 586, 512 

  \bibitem[2005]{Bur05} Burgasser, A. J., Reid, I. N., Leggett, S.K., et~al.
                        2005, ApJ, 634, 177 

  \bibitem[2006a]{Bur06} Burgasser, A. J., Geballe, T. R., Leggett, S.K., et~al.
                         2006a, ApJ, 637, 1067 

  \bibitem[2006b]{Bur06b} Burgasser, A. J., Kirkpatrick, D. J., Cruz, K. L., et~al.
                         2006b, ApJS, 166, 585 
                         
\bibitem[2006]{Chi06} Chiu, K., Fan, X., Leggett, S. K., et~al.
                         2006, AJ, 131, 2722
                         
  \bibitem[2002a]{Cla02b} Clarke, F. J., Tinney, C. G. \& Covey, K. R.
                          2002a, MNRAS, 332, 361 

  \bibitem[2002b]{Cla02a} Clarke, F. J., Oppenheimer, B. R. \& Tinney, C. G.
                        2002b, MNRAS, 335, 1158 

  \bibitem[2003]{Cla03} Clarke, F. J., Tinney, C. G. \& Hodgkin, S. T.
                        2003, MNRAS, 341, 239 

  \bibitem[2003]{Cus03} Cushing, M. C., Rayner, J. T., Davis, S. P. \& Vacca, W. D. 
	                2003, ApJ, 582, 1066

  \bibitem[2004]{Cus04} Cushing, M. C., Vacca, W. D. \& Rayner, J. T.
                        2004, PASP, 116 362

{\bibitem[2005]{Cus05} Cushing, M. C., Rayner, J. T., \& Vacca, W. D. 
                        2005, ApJ, 623, 1115}

\bibitem[2008]{Cus08} Cushing, M. C., Marley, M. S., Saumon, D. ~et~al.
                        2008, ApJ, {\it in press} ({\tt astro-ph/0711.0801})
                        
\bibitem[2002]{Dah02} Dahn, C. C., Harris, H. C., Vrba, F. J., et~al.
                        2002, AJ, 124, 1170 

  \bibitem[2003]{Eno03} Enoch, M. L., Brown, M. E. and Burgasser, A. J.
                        2003, AJ, 126, 1006

  \bibitem[2002]{Geb02} Geballe, T. R., Knapp, G. R., Leggett, S. K., et~al.
                        2002, ApJ, 564, 466

  \bibitem[2000]{Gel00} Gelino, C. R. \& Marley, M. S.
                        2000, in ASP Conf. Ser. 212, From Giant Planet to Cool Stars, 
                        Edited by C.~A.~Griffith \&~M.~S.~Marley (San Francisco: ASP), 322 

  \bibitem[2002]{Gel02} Gelino, C. R., Marley, M. S., Holtzman, J. A., et~al.
                        2002, ApJ, 577, 433

  \bibitem[2003]{Gol03} Goldman, B., for the {\sc Clouds} collaboration,
                        2003, Brown Dwarfs, IAU Symposium \#211. Edited by E.~Mart\'\i n. San Francisco: PASP, 461

  \bibitem[2004]{Gol04} Golimowski, D. A., Leggett, S. K., Marley, M. S., et~al.
                        2004, AJ, 127, 3516

  \bibitem[2002]{Hal02} Hall, P. B.
                        2002, ApJ, 564, L89

  \bibitem[1999]{Kir99} Kirkpatrick, J. D., Reid, I. N., Liebert, J., et~al.
                        1999, ApJ, 519, 802 

  \bibitem[2000]{Kir00} Kirkpatrick, J. D., Reid, I. N., Liebert, J., et~al.
                        2000, AJ, 120, 447
                        
  \bibitem[2001]{Kir01} Kirkpatrick, J. D., Dahn, C. C., Monet, D. G.,
                        2001, AJ, 121, 3235

  \bibitem[2004]{Kna04} Knapp, G. R., Leggett, S. K., Fan, X., et~al.
                        2004, AJ, 127, 3553
                        
  \bibitem[2003]{Koe03} Koen, C.
                        2003, MNRAS, 346, 473
                        
  \bibitem[2000]{Leg00} Leggett, S. K., Geballe, T. R., Fan, X., et~al.
                        2000, ApJ, 536, L35
                        
  \bibitem[2003]{Lie03} Liebert, J., Kirkpatrick, J. D., Cruz, K. L., et~al.
                        2003, AJ, 125, 343
 
   \bibitem[2006]{Liu06} Liu, M. C., Leggett, S. K., Golimowski, D. A., et~al.
                        2006, ApJ, 647, 1393 

  \bibitem[2003]{McL03} McLean, I. S., Mc~Govern, M. R., Burgasser, A. J., et~al.
                        2003, ApJ, 596, 561

  \bibitem[2002]{Mar02} Marley, M. S., Seager, S., Saumon, D., et~al.
                        2002, ApJ, 568, 335

  \bibitem[1999]{Mar99} Mart\'\i n, E. L., Delfosse, X., Basri, G., et~al.
                        1999, AJ, 118, 2466

  \bibitem[2003]{Moh03} Mohanty, S. \& Basri, G.
                        2003, ApJ, 583, 451 

\bibitem[2006]{Mor06} Morales-Calder{\'o}n, M., Stauffer, J. R., Kirkpatrick, J. D., et~al.
                         2006, \apj, 653, 1454 

  \bibitem[2000]{Nak00} Nakajima, T., Tsuji, T., Maihara, T., et~al.
                        2000, PASJ, 52, 87 

  \bibitem[2003]{Ray03} Rayner, J. T., Toomey, D. W., Onaka, P. M., et~al.
                        2003, PASP, 115, 362
                        
  \bibitem[2001]{Rei01} Reid, I. N., Gizis, J. E., Kirkpatrick, J. D., \& Koerner, D.
                        2001, AJ, 121, 489

  \bibitem[2004]{Roe04} Roelling, T. L., VanÊCleve,ÊJ.ÊE., Sloan,ÊG.ÊC., et~al.~
                        2004, ApJS, 154, 418
  
  \bibitem[1999]{Str99} Strauss, M. A.; Fan, X.; Gunn, J. E., et~al.
                         1999, ApJ, 522, L61

  \bibitem[1999]{Tin99} Tinney, C. G. \& Tolley, A. J.
                        1999, MNRAS, 304, 119 

  \bibitem[2003]{Tin03} Tinney, C. G., Burgasser, A. J. \& Kirkpatrick, J. D.
                        2003, AJ, 126, 975 

  \bibitem[2004]{Tsu04} Tsuji, T., Nakajima, T. \& Yanagisawa, K.
                        2004, ApJ, 607, 511

  \bibitem[2003]{Vac03} Vacca, W. D., Cushing, M. C. \& Rayner, J. T.
                        2003, PASP 155, 389

  \bibitem[2004]{Vrb04} Vrba, F. J., Henden, A. A., Luginbuhl, H. H., et~al.
                        2004, AJ, 127, 2948
                        
  \bibitem[1977]{Win77} Wing, , Cohen, , \& Brault, ,
                        1977, 
                        
  \bibitem[2006]{ZO06} Zapatero Osorio, M.~R., Mart{\' \i}n, E. L., Bouy, H., et~al.
                        2006, ApJ, 647, 1405

\end{thebibliography}
\end{document}